\documentclass[journal,twoside,web]{ieeecolor}
\usepackage{generic}
\usepackage{cite}
\usepackage{amsmath,amssymb,amsfonts}
\usepackage{algorithmic}
\usepackage{graphicx}
\usepackage{textcomp}

\newcommand{\tabincell}[2]{\begin{tabular}{@{}#1@{}}#2\end{tabular}}
\newcommand{\be}{\begin{equation}}
\newcommand{\ee}{\end{equation}}
\newcommand{\bea}{\begin{eqnarray}}
\newcommand{\eea}{\end{eqnarray}}
\def\QEDclosed{\mbox{\rule[0pt]{1.3ex}{1.3ex}}}
\def\QED{\QEDclosed}
\def\endproof{\hspace*{\fill}~\QED\par\endtrivlist\unskip}

\newtheorem{theorem}{Theorem}
\newtheorem{remark}{Remark}
\newtheorem{lemma}{Lemma}
\newtheorem{definition}{Definition}

\usepackage{tabularx}
\usepackage{dsfont}
\usepackage{bbding}

\usepackage{booktabs}

\usepackage{subfigure}

\def\BibTeX{{\rm B\kern-.05em{\sc i\kern-.025em b}\kern-.08em
    T\kern-.1667em\lower.7ex\hbox{E}\kern-.125emX}}
\begin{document}
\title{Attack Detection for Networked Control Systems Using Event-Triggered Dynamic Watermarking}
\author{Dajun Du, Changda Zhang, Xue Li, Minrui Fei, and Huiyu Zhou
\thanks{The work of D. Du, C. Zhang, X. Li, and M. Fei was supported in part by the National Science Foundation of China under Grant Nos. 92067106, 61773253, 61803252, and 61833011, 111 Project under Grant No. D18003, and Project of Science and Technology Commission of Shanghai
Municipality under Grant Nos. 20JC1414000, 19510750300, 21190780300. \emph{(Corresponding authors: Changda Zhang; Xue Li.)}}
\thanks{D. Du, C. Zhang, X. Li, and M. Fei are with Shanghai Key Laboratory of Power Station Automation Technology, School of Mechatronic Engineering
and Automation, Shanghai University, Shanghai 200444, China (e-mail: ddj@i.shu.edu.cn; changdazhang@shu.edu.cn; lixue@i.shu.edu.cn; mrfei@staff.shu.edu.cn). }
\thanks{H. Zhou is with School of Computing and Mathematical Sciences, University of Leicester, Leicester LE1 7RH, U.K. (e-mail: hz143@leicester.ac.uk).}}

\maketitle

\begin{abstract}
Dynamic watermarking schemes can enhance the cyber attack detection capability of networked control systems (NCSs). This paper presents a linear event-triggered solution to conventional dynamic watermarking (CDW) schemes. Firstly, the limitations of CDW schemes for event-triggered state estimation based NCSs are investigated. Secondly, a new event-triggered dynamic watermarking (ETDW) scheme is designed by treating watermarking as symmetric key encryption, based on the limit convergence theorem in probability. Its security property against the generalized replay attacks (GRAs) is also discussed in the form of bounded asymptotic attack power. Thirdly, finite sample ETDW tests are designed with matrix concentration inequalities. Finally, experimental results of a networked inverted pendulum system demonstrate the validity of our proposed scheme.
\end{abstract}


\section{Introduction}
\label{sec:introduction}
\IEEEPARstart{M}{odern} networked control systems (NCSs) integrate communication networks and smart sensors with physical plants and digital controllers (including digital filters) in an effort to achieve satisfactory efficiency and productivity over traditional control systems \cite{NCS1,NCS2,NCS3}.

However, the wide introduction and usage of communication networks (especially wireless networks \cite{SAT}) may bring the following two essential problems. (1) \emph{NCSs security.} Compared with self-enclosed physical plants and digital controllers, the cyber space of networks is open to real world, which makes networks easy to access by malicious users. Recent years have witnessed serious cyber threats to NCSs such as the attack on U.S. fuel pipelines \cite{USC} and Stuxnet worm \cite{STU} on Iranian nuclear facilities. (2) \emph{Energy scarcity.} Specifically, when distributed sensors for measuring plant performance communicates with digital controllers via a wireless network and are battery-powered, the energy supply of sensors becomes limited and it is usually not realistic to achieve frequent battery replacement. It is widely known \cite{EHS} that compared with data sampling, data transmission for sensors equipped with radio modules usually consumes more energy; \emph{e.g.}, the power consumption of temperature sampling of sensor STCN75 produced by STM is \emph{0.4 mW}, and the power consumption of data transmission of radio module CC2420 produced by Texas Instruments is \emph{35 mW (at 0dBm)}. In this context, it is not surprising that the critical questions on NCSs security and energy scarcity have attracted wide attention.

NCSs security can be greatly guaranteed by addressing the problems of attack modelling and detection. A popular attack is deception attack, which has a major type as replay attacks \cite{STU}. A generalization of replay attacks is termed as generalized replay attacks (GRAs) \cite{GRA,GRA2}, which is with simplicity of implementation and has been applied in real-world attacks. It has been found \cite{OLA,DWM1} that both replay attacks and GRAs have an ability of bypassing passive detection methods (\emph{e.g.}, the $\chi^2$ test), meanwhile destroying stability of systems with unstable open-loop dynamics. To improve the attack detection capability of passive detection methods, active detection methods have been widely investigated.

Dynamic watermarking is one particular active detection method, which mainly includes CDW scheme \cite{DWM1,DWM2,DWL} and new dynamic watermarking (NDW) scheme \cite{NCS3,SCN}. For CDW scheme, the control or actuator signals are encrypted by injecting a watermarking signal generally with Gaussian distributions. For NDW scheme, the before-transmission system outputs are encrypted by injecting a watermarking signal, and then the after-transmission system outputs are decrypted by the same watermarking signal. For CDW and NDW scheme, some hypothesis tests (\emph{e.g.}, $\chi^2$ test \cite{DWM1}, KL divergence based test \cite{NCS3} or consistent tests \cite{DWM2,DWL,SCN}) are used to detect cyber attacks. Generally, the consistent tests are designed to be asymptotic (\emph{i.e.}, with infinite window sizes) at the beginning, and then are transformed to be statistical (\emph{e.g.}, with finite window sizes \cite{GRA,DWM4} or finite samples \cite{FSDW}) for practical use in real world.

Existing dynamic watermarking schemes have the following two features. (1) Under no attack, nonzero system performance loss is introduced by watermarking signals for CDW scheme, while NDW scheme can guarantee zero system performance loss. (2) For linear systems with \emph{time-triggered commutation} (TTC, \emph{i.e.}, periodic sampling and data transmission), the replay attacks detection capability increases with watermarking intensity increasing for CDW scheme \cite{DWM1}; the detection of attacks is guaranteed by the asymptotic CDW \cite{DWM2} and NDW schemes \cite{SCN}; the finite failure on GRAs detection is guaranteed by the finite sample CDW scheme \cite{FSDW}; the detection of GRAs is guaranteed by the asymptotic time-varying CDW scheme \cite{GRA}. In a word, existing dynamic watermarking schemes have only focused on TTC. While TTC can be well-used for stable operation and superior performance of systems, the use of TTC may be limited by energy scarcity.

Energy scarcity compels some energy-recharging solutions \cite{EES} (\emph{e.g.}, energy harvesting and transferring) or energy-saving solutions to be extensively used, where specially the energy-saving solutions are indispensable when there are no other resources (\emph{e.g.}, wind and solar power) for recharging energy. \emph{Event-triggered communication} (ETC)\cite{NCS3,ETC1,ETC2} is one popular energy-saving solution, where the sampled data will be sent through networks immediately when a predefined event-triggered condition is violated at a certain time instant. Specially, if the state of physical plants is partially available, the event-triggered state estimation (ETSE) \cite{ETSE1,ETSE2} developed from the standard Kalman filter is commonly used to obtain the state estimate. The primary advantage of ETC compared with TTC is saving energy, meanwhile maintaining the comparable performance of systems \cite{NCS3}. While ETC provides the above advantages, existing dynamic watermarking schemes may not always be applicable for ETC.

Motivated by the above observations, the following challenges will be addressed:
\begin{enumerate}
  \item[(1)]
  What are the limitations of CDW scheme for ETSE-based NCSs?
  \item[(2)]
  How to design a new dynamic watermarking scheme for ETSE-based NCSs? What is the security property of such scheme?
  \item[(3)]
  For the new dynamic watermarking scheme, how to design new statistical tests available for real-world ETSE-based NCSs?
\end{enumerate}

\begin{table}[!t]
\centering
\caption{Comparison between the Existing Dynamic Watermarking Schemes and Our Proposed Scheme.}
\label{Tabsum}
\begin{tabular}{llcll}
\toprule
  Refs.$^1$ & DWF$^2$ & ECS$^3$/ESO$^4$/SPL$^5$ & LSF$^6$ & TCM$^7$ \\
\midrule
  \cite{GRA}  & Asym.$^8$ \& FWS$^9$ & \CheckmarkBold/\XSolidBrush/\CheckmarkBold & LTV$^{10}$ & TTC$^{11}$  \\
  \cite{DWM1}  & FWS                  & \CheckmarkBold/\XSolidBrush/\CheckmarkBold & LTI$^{12}$ & TTC  \\
  \cite{DWM2}  & Asym.                & \CheckmarkBold/\XSolidBrush/\CheckmarkBold & LTI        & TTC  \\
  \cite{DWL}  & Asym. \& FWS         & \CheckmarkBold/\XSolidBrush/\CheckmarkBold & LTI        & TTC  \\
  \cite{DWM4} & FWS                  & \CheckmarkBold/\XSolidBrush/\CheckmarkBold & LTI        & TTC  \\
  \cite{FSDW}  & Asym. \& FIS$^{13}$  & \CheckmarkBold/\XSolidBrush/\CheckmarkBold & LTI        & TTC  \\
  \cite{NCS3}  & FWS                  & \XSolidBrush/\CheckmarkBold/\XSolidBrush   & LTI        & TTC  \\
  \cite{SCN}  & Asym. \& FWS         & \XSolidBrush/\CheckmarkBold/\XSolidBrush   & LTI        & TTC  \\
  $\dag^{14}$  & Asym. \& FIS        & \XSolidBrush/\CheckmarkBold/\XSolidBrush   & LTI        & \textbf{ETC}$^{15}$  \\
\bottomrule
\multicolumn{5}{l}{$^1$References. $^2$Dynamic watermarking form. }\\
\multicolumn{5}{l}{$^3$Encryption of control signal. $^4$Encryption of system outputs.}\\
\multicolumn{5}{l}{$^5$System performance loss. $^6$Linear system form.}\\
\multicolumn{5}{l}{$^7$Triggering communication modes. $^8$Asympototic. $^9$Finite window size.}\\
\multicolumn{5}{l}{$^{10}$Linear time-varying. $^{11}$Time-triggering communication.}\\
\multicolumn{5}{l}{$^{12}$Linear time-invariant. $^{13}$Finite sample.}\\
\multicolumn{5}{l}{$^{14}$This paper. $^{15}$Event-triggered communication.}\\
\end{tabular}
\end{table}
To deal with these challenges, this paper extends CDW scheme into a new event-triggered dynamic watermarking (ETDW) scheme. Compared with the existing results in the literatures, comparative analysis is listed in Table~\ref{Tabsum}. It can be clearly found that the existing results have only focused on dynamic watermarking scheme for TTC, but the proposed new ETDW scheme pays its attention to dynamic watermarking scheme for ETC. The main contributions of this paper are summarized as follows:
\begin{enumerate}
  \item[(1)]
  The limitations of CDW scheme for ETSE-based NCSs are revealed, where there are system performance loss from watermarking signals and event-triggered covariance of signals used for attack detection;
  \item[(2)]
  A new ETDW scheme is designed by treating watermarking as symmetric key encryption, based on limit convergence theorem in probability. Furthermore, the security property of such scheme is proven that the asymptotic power of undetected GRAs is guaranteed to be not more than the power of attack-free residuals.
  \item[(3)]
  Two new finite sample ETDW tests are designed by using matrix concentration inequalities. Furthermore, the attack detection capability of such tests is proved with finite false alarm under no attack and finite failure on GRAs detection.
\end{enumerate}

The rest of this paper is organized as follows. Section~\uppercase\expandafter{\romannumeral+2} is problem formulation, where ETSE-based systems with CDW scheme under the GRAs and limitations of CDW scheme are analysed. Section~\uppercase\expandafter{\romannumeral+3} presents a new ETDW scheme, where the design, security property analysis, and performance analysis of our proposed approach are presented. Experimental results for a networked inverted pendulum system are given in Section~\uppercase\expandafter{\romannumeral+4}, followed by the conclusion in Section~\uppercase\expandafter{\romannumeral+5}.

\emph{Notation:} The Euclidean norm of a vector $x$ is denoted as $\|{x}\|$. The spectral norm and spectral radius of a matrix $X$ are denoted as $\|{X}\|$ and $\rho(X)$, respectively. Multivariate Gaussian distribution with mean $\mu$ and covariance $\mathcal{E}$ is denoted $\mathcal{N}(\mu,\mathcal{E})$. The expectation of a random variable $a$ conditional on the variable $b$ is denoted $\mathds{E}[a|b]$. Given two events $E_1$ and $E_2$, the probability of $E_1$ conditional on $E_2$ is denoted $\mathds{P}(E_1|E_2)$ and the inverse event of $E_1$ is denoted $\neg {E_1}$. Table~A.\uppercase\expandafter{\romannumeral+2} of Section~\uppercase\expandafter{\romannumeral+1} in the supplementary materials summarizes the notations most frequently used throughout the rest of the paper.

\section{Problem Formulation}
\subsection{ETSE-Based NCSs with CDW Scheme}
\begin{figure}[!t]
  \centering
  \includegraphics[width=0.485\textwidth]{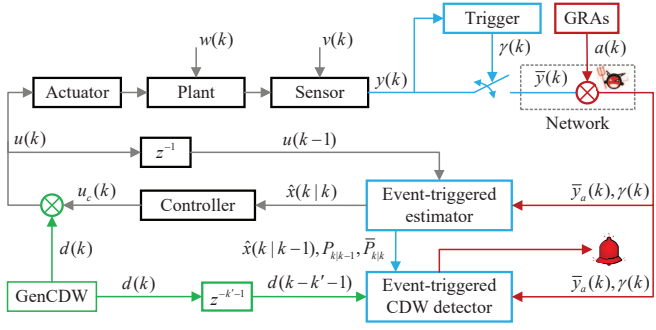}
  \caption{Framework of ETSE-based NCSs with CDW scheme.}
  \label{fig1}
\end{figure}
The general framework of ETSE-based NCSs with CDW scheme is shown in Fig.~\ref{fig1}. The system output $y(k)$ is firstly measured by the sensor periodically. After having received $y(k)$, the trigger generates a binary triggering signal $\gamma(k)$ by the preset triggering condition and then accordingly decides whether or not $y(k)$ is sent to the network; as a consequence, $y(k)$ becomes $\bar y(k)$. Then, $\bar y(k)$, $\gamma(k)$ will be transmitted to the event-triggered estimator and the CDW detector via the network, which may be attacked and become $\bar y_a(k)$, $\gamma(k)$. Using received $\bar y_a(k)$, $\gamma(k)$ and the control input $u(k-1)$, the event-triggered estimator calculates the \emph{a priori} state estimate $\hat x(k|k-1)$, \emph{a posteriori} state estimate $\hat x(k|k)$, prediction error covariance $P_{k|k-1}$ and upper bound $\bar P_{k|k}$ of estimation error covariance. Furthermore, using $\hat x(k|k-1)$, $P_{k|k-1}$, $\bar P_{k|k}$ and watermarking signal $d(k-k'-1)$ generated by GenCDW, the event-triggered CDW detector judges whether or not an attack takes place; if yes, the alarm will sound, otherwise the alarm keeps silence. Furthermore, using $\hat x(k|k)$, the controller calculates $u_c(k)$, which is encrypted by $d(k)$ and becomes $u(k)$. Finally, the actuator applies $u(k)$ to stabilizing the plant.
\begin{remark}
It is not necessary to transmit $\gamma(k)$ in real world. The event-triggered estimator and the CDW detector have the ability to judge whether or not a data packet is received, \emph{e.g.}, by the mechanism of TCP-ACK transmission \cite{OPU}.
\end{remark}

To analyse the limitations of CDW scheme for ETSE-based NCSs, we consider a discrete-time linear time-invariant plant
\begin{align}
  x(k + 1) &= Ax(k) + B u(k) + w(k), \hfill \label{eq21}\\
  y(k) &= C x(k) + v(k) \hfill \label{eq22}
\end{align}
where $x(k) \in \mathbb{R}^{n_x}$, $u(k) \in \mathbb{R}^{n_u}$ and $y(k) \in \mathbb{R}^{n_y}$ are the system state, control input and system output at $k$-th sampling instant, respectively; the process noise $w(k)$ and measurement noise $v(k)$ are mutually independent and take the distribution form $w(k) \sim \mathcal{N}(0,\mathcal{E}_w)$, $v(k) \sim \mathcal{N}(0,\mathcal{E}_v)$. $A$, $B$ and $C$ are constant matrices with appropriate dimensions.

\subsubsection{ETC with Send-on-Delta Condition}
The send-on-delta condition for ETC is used to send $y(k)$. To formulate the send-on-delta condition, a triggering signal $\gamma(k)$ is defined by
\be
\gamma (k) := \left\{ \begin{gathered}
  1,\ if\ {\epsilon ^T}(k)\epsilon (k) > \delta  \hfill \\
  0,\ otherwise \hfill \\
\end{gathered}  \right.
\label{eq23}
\ee
where $\epsilon (k): = y(k) - y(\tau_k)$ is the difference between $y(k)$ and the previously transmitted measurement $y(\tau_{k})$ (and $\tau_{k} < k$); $\delta >0$ is the user-defined event-triggered threshold. Note that if and only if $\gamma(k)=1$, then $y(k)$ will be sent via the network. Consequently, the network's input becomes
\be
\bar y(k) = y(k) - \left( {1 - \gamma (k)} \right)\epsilon (k).
\label{eq24}
\ee

\subsubsection{The GRAs}
$\bar y(k)$ could be compromised by cyber attacks from the network. Note that if and only if there is no any attack, then $\bar y_a(k) = \bar y(k)$. When there is a persistent event-triggered version of GRAs \cite{GRA}, $\bar y_a(k)$ can be formulated by
\begin{align}
  {{\bar y}_a}(k) &= \bar y(k) + a(k), \hfill \label{eq25}\\
  a(k) &= \gamma (k)\left( {s\bar y(k) + C{x_a}(k) + v_a(k)}\right),  \hfill \label{eq26}\\
  {x_a}(k+1) &= {A_a}{x_a}(k) \hfill \label{eq27}
\end{align}
where $a(k)$ is the false data with respect to $\gamma(k)$ and generated by the hidden Markov model (\ref{eq26}) and (\ref{eq27}); $s \in \mathbb{R}$ is called the attack scaling factor; $x_a(k) \in \mathbb{R}^{n_x}$ is the hidden state of GRAs; the noise ${v_a}(k)$ takes the distribution of $v_a(k) \sim \mathcal{N}(0,\mathcal{E} _{v_a})$, and $v_a(k)$ is mutually independent with $v(k)$, $w(k)$; $A_a$ is Schur stable, \emph{i.e.}, $\rho(A_a)<1$.

To further quantify the persistent additional distortion from GRAs, we consider the following Definition~\ref{D1}.
\begin{definition}
\label{D1}
The \emph{asymptotic attack power} is defined as, cf. \cite{DWL},
\be
\mathop {{\text{\rm as-lim}}}\limits_{i \to \infty } \frac{1}{i}\sum\nolimits_{k = 1}^{i} {{a^T}(k)a(k)}
\label{eq28}
\ee
where $\text{\rm as-lim}$ represents the almost sure limit, or, cf. \cite{GRA},
\be
\mathop {{\text{\rm p-lim}}}\limits_{i \to \infty } \frac{1}{i}\sum\nolimits_{k = 1}^{i} {{a^T}(k)a(k)}
\label{eq29}
\ee
where $\text{\rm p-lim}$ represents limit in probability.
\end{definition}

\subsubsection{ETSE with Send-on-Delta condition}
Using $\bar y_a(k)$, $\gamma(k)$ and $u(k-1)$, an event-triggered estimator can be designed to estimate system states for calculating control input, \emph{i.e.},
\begin{align}
  \hat x(k|k - 1) &= A\hat x(k - 1|k - 1) + Bu(k - 1), \hfill \label{eq210}\\
  \hat x(k|k) &= \hat x(k|k - 1) + L(k,\gamma,\delta)r(k) \hfill \label{eq211}
\end{align}
where $r(k):={\bar y_a(k) - C\hat x(k|k - 1)}$ is the residual (or innovation); $L(k,\gamma,\delta)$ is the event-triggered estimator gain and can be designed according to Theorem~A.1 of Section~II.A in the supplementary materials.

\subsubsection{Controller and CDW Scheme}
Using $\hat x(k|k)$ generated by the above event-triggered estimator, the controller output can be calculated by minimizing linear quadratic Gaussian performance $J$ \cite{DWM1}, and the optimal solution is
\be
{u_c}(k) = K\hat x(k|k)
\label{eq216}
\ee
where $K =  - {\left( {{B^T}SB + R} \right)^{ - 1}}{B^T}SA$ is the controller gain; $S$ is the unique positive definite solution of algebraic Riccati equation $S = {A^T}SA + Q - {A^T}SB{\left( {{B^T}SB + R} \right)^{ - 1}}{B^T}SA$; $Q$ and $R$ are the designed constant matrices. Furthermore, to detect cyber attacks, ${u_c}(k)$ is encrypted by injecting $d(k)$, \emph{i.e.},
\be
u(k) = {u_c}(k) + d(k)
\label{eq217}
\ee
where $d(k) \sim \mathcal{N}(0,\mathcal{E}_d)$ is independent of $u_c(k)$, and $\mathcal{E}_d$ is full rank.

The ETSE-based NCSs with CDW scheme have been well partially established as (\ref{eq21})--(\ref{eq27}) and (\ref{eq210})--(\ref{eq217}), where there are still attack detection formulas remaining to be designed. Then, two asymptotic CDW tests have been designed \cite{DWL} and the corresponding security property restricting the attack power (\ref{eq28}) of undetected attacks to zero is given in Theorem~A.2 of Section~II.C in the supplementary materials.

\begin{remark}
The above event-triggered estimator (\ref{eq210}), (\ref{eq211}) and controller (\ref{eq217}) can theoretically satisfy the state limitation, \emph{i.e.}, if $\rho \left( {A+BK} \right) < 1$ and $\rho \left( {(\mathds{I}-L(k,\gamma,\delta)C)A} \right) < 1$, then $\exists {\varsigma} < \infty$, it follows that ${\lim _{k \to \infty }}\mathds{E}\left[ {\left\| {x(k)} \right\|} \right] < {\varsigma}$, where the proof is given in Section~II.B of the supplementary materials. Moreover, in the experiments of Section~\uppercase\expandafter{\romannumeral+4}, appropriate $Q$ and $R$ are selected to guarantee $\rho \left( {A+BK} \right) < 1$, and proper $\delta$ is also selected to guarantee $\rho \left( {(\mathds{I}-L(k,\gamma,\delta)C)A} \right) < 1$, meanwhile $Q$, $R$ and $\delta$ are carefully regulated so that the system states do not exceed the limitation.
\end{remark}

\begin{figure*}[!t]
  \centering
  \includegraphics[width=0.7\textwidth]{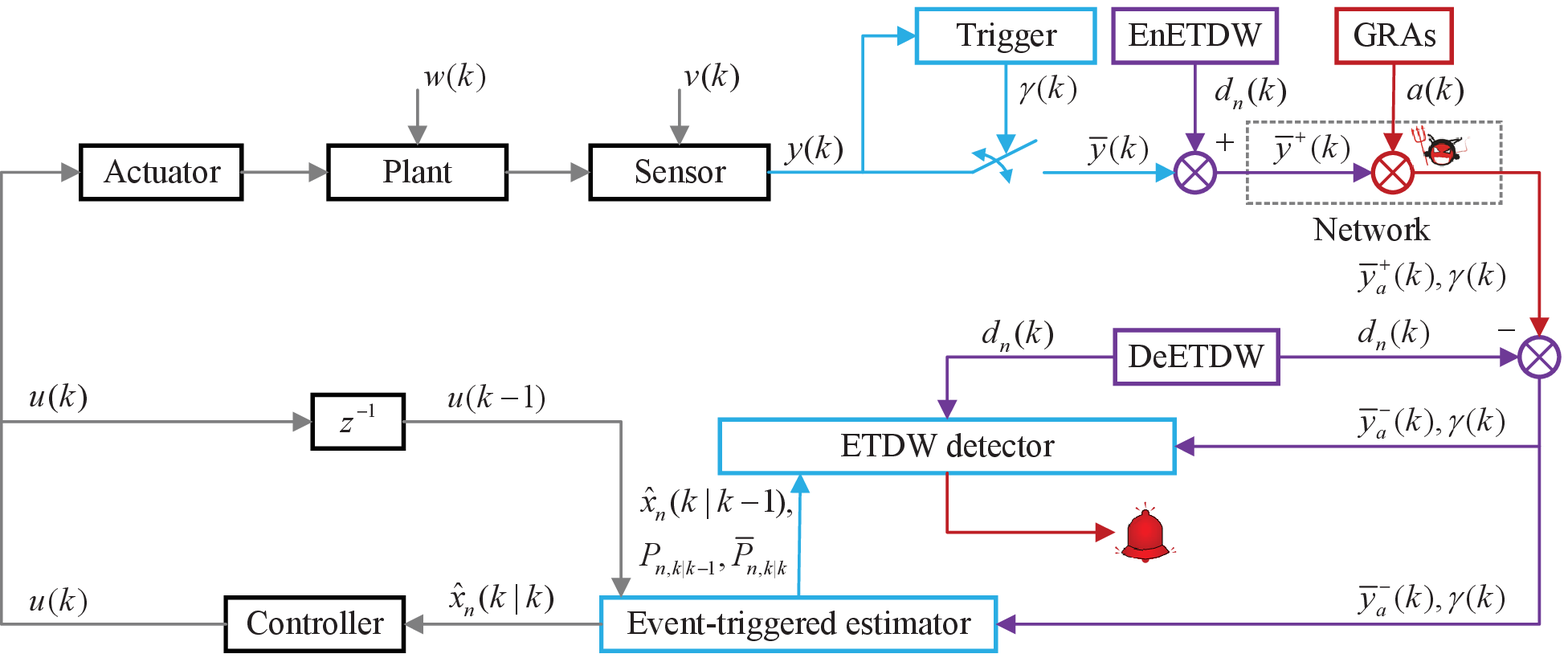}
  \caption{Framework of ETSE-based NCSs with new ETDW scheme.}
  \label{fig2}
\end{figure*}
\begin{remark}
The infinite limits $i \to \infty$ in the asymptotic CDW tests (A.14) of Section~II.C in the supplementary materials are not well suited for real-time attack detection. To deal with the problem, the finite sample CDW tests \cite{FSDW} have been transformed from the asymptotic CDW tests (A.14) of Section~II.C in the supplementary materials, where two events are defined by
\begin{subequations}
\label{eq219}
\begin{align}
{\Xi _{1,i}} &:= \left\{ {\left\| {\frac{1}{i}{\mathcal{D}_i}} \right\| \geqslant {\vartheta _{1,i}}} \right\},     \hfill \label{eq219a}\\
{\Xi _{2,i}} &:= \left\{ {\left\| {\frac{1}{i}\left( {{\mathcal{R}_i} - i\mathcal{E}_r^f} \right)} \right\| \geqslant {\vartheta _{2,i}}} \right\}   \hfill \label{eq219b}
\end{align}
\end{subequations}
where ${\vartheta _{1,i} }$ and $\vartheta _{2,i}$ are time-varying threshold functions. Furthermore, define ${\Xi _i}: = \left\{ {{\Xi _{1,i}} \cup {\Xi _{2,i}}} \right\}$, and if ${\Xi _i}$ is true, then the attack will be alarmed at the $i$-th sampling instant, otherwise the alarm keeps silence.
\end{remark}

\subsection{Limitations of CDW Scheme for ETSE-Based NCSs}
The above discussion is for ETSE-based NCSs with CDW scheme. However, due to the introduction of watermarking signals and ETSE, there are two limitations of CDW scheme for ETSE-based NCSs.

\emph{Limitation 1--System Performance Loss From Watermarking:} Considering the system (\ref{eq21})--(\ref{eq27}) and (\ref{eq210})--(\ref{eq217}) under no attacks or trigger (\emph{i.e.}, $a(k)=0$, $\gamma(k)=1$, $\forall k \geqslant 0$), then the system performance loss is, cf. \cite[Th. 3]{DWM1},
\be
\Delta {J} = tr\left( {({B^T}SB + R){\mathcal{E} _d}} \right)
\label{eq220}
\ee
where $S$ and $R$ have been given in (\ref{eq216}).

\emph{Limitation 2--Event-Triggered Covariance:} Considering the system (\ref{eq21})--(\ref{eq27}) and (\ref{eq210})--(\ref{eq217}) under no attack (\emph{i.e.}, $a(k)=0, \forall k \geqslant 0$), the cross covariance of $r(k)$ and $d(k-1)$, and the auto covariance of $r(k)$ satisfy
\begin{subequations}
\label{eq221}
\begin{align}
  \mathds{E}\left[ {\left. {r(k){d^T}(k - 1)} \right|\gamma (k)} \right] &=  \left( {\gamma (k)-1} \right)CB{\mathcal{E} _d}, \hfill \label{eq221a}\\
  \mathds{E}\left[ {\left. {r(k){r^T}(k)} \right|\gamma (k)} \right] &\leqslant {\Psi _\beta }\left({k,\gamma,\delta}\right) \hfill \label{eq221b}
\end{align}
\end{subequations}
where ${\Psi _\beta }\left({k,\gamma,\delta}\right)$ is the same as (A.3) of Section~II.A in the supplementary materials. The proof is given in Section~\uppercase\expandafter{\romannumeral+2}.D of the supplementary materials.
\begin{remark}
From limitations 1 and 2, it can be clearly seen that there are three main problems of CDW scheme for ETSE-based NCSs to be solved: 1) bigger $\mathcal{E}_d$ brings more $\Delta {J}$; 2) the asymptotic CDW tests (A.14) of Section~II.C in the supplementary materials may not be used; and 3) the finite sample CDW tests (\ref{eq219}) may also not be used. To overcome limitations 1 and 2, it is necessary to develop a new attack detection scheme.
\end{remark}

\section{New Event-Triggered Dynamic Watermarking}
The above has presented the framework of ETSE-based NCSs with CDW scheme, and analysed the corresponding two limitations. To cope with the limitations, a new ETDW scheme is designed and analysed from \emph{asymptotic} form to \emph{finite sample} form. As a comparison, a candidate ETDW scheme is designed and analysed in Section \uppercase\expandafter{\romannumeral+3}.A of the supplementary materials.

\subsection{Design and Security Property Analysis of New ETDW Scheme for ETSE-Based NCSs}
The designed framework of ETSE-based NCSs with new ETDW scheme is shown in Fig.~\ref{fig2}. Firstly, $y(k)$ is measured by the sensor periodically. After having received $y(k)$, the trigger generates $\gamma(k)$ by (\ref{eq23}) and then accordingly decides whether or not $y(k)$ is sent to the network; as a result, $y(k)$ becomes $\bar y(k)$. Secondly, $\bar y(k)$ is encrypted by injecting $d_n(k)$ that is generated by EnENDW, becoming $\bar y^{+}(k)$. Thirdly, $\bar y^+(k)$, $\gamma(k)$ are transmitted to the event-triggered estimator and the ETDW detector via the network, which may be attacked and become $\bar y^+_a(k)$, $\gamma(k)$. Then, $\bar y_a^+(k)$ is decrypted by subtracting $d_n(k)$ that is generated by DeENDW, becoming $\bar y_a^{-}(k)$. Using received $\bar y _a^-(k)$, $\gamma(k)$ and $u(k-1)$, the event-triggered estimator calculates $\hat x_n(k|k-1)$, $\hat x_n(k|k)$, $P_{n,k|k-1}$ and $\bar P_{n,k|k}$. Furthermore, using $\hat x_n(k|k-1)$, $P_{n,k|k-1}$, $\bar P_{n,k|k}$ and $d_n(k)$, the ETDW detector evaluates whether or not an attack takes place; if yes, the alarm will sound, otherwise the alarm keeps silence. Furthermore, using $\hat x_n(k|k)$, the controller calculates $u(k)$, which is applied by the actuator to stabilizing the plant.

\subsubsection{Design of Watermarking as Symmetric Key Encryption, Event-Triggered Estimator and Controller}
Consider the plant and trigger (\ref{eq21})--(\ref{eq24}). To monitor the information integrity and protect the information confidentiality, $\bar y(k)$ is encrypted by $d_n(k)$, becoming
\be
{\bar y^ + }(k) = \bar y(k) + d_n(k)
\label{eq3A1}
\ee
where $d_n(k) \sim \mathcal{N}(0,\mathcal{E}_{d_n})$ is independent of $\bar y(k)$, and $\mathcal{E}_{d_n}$ is with full rank. Under the GRAs, ${\bar y^ + }(k)$ becomes
\be
\begin{aligned}
\bar y_a^ + (k) &= {{\bar y}^ + }(k) + a(k), \hfill \\
a(k) &= \gamma (k)\left( {s{{\bar y}^ + }(k) + C{x_a}(k) + {v_a}(k)} \right), \hfill \\
x_a(k+1) &= A_a x_a(k). \hfill
\end{aligned}
\label{eq3A2}
\ee
To prevent the watermarking signal exciting the system operation, $\bar y_a^ + (k)$ is decrypted by $d_n(k)$, becoming
\be
\bar y_a^ - (k) = \bar y_a^ + (k) - d_n(k).
\label{eq3A3}
\ee
For now, the mechanism of watermarking as symmetric key encryption (\ref{eq3A1}) and (\ref{eq3A3}) is completed, which has been developed in \cite{SCN}, and guarantees attack-free and trigger-free
\be
\Delta{J} = 0.
\label{eq3A4}
\ee

Using $\bar y_a^ - (k)$, the event-triggered estimator is designed as
\begin{align}
\label{eq3A5} \hat x_n(k|k-1) &= A \hat x_n(k-1|k-1) + B u(k-1), \hfill \\
\label{eq3A6} \hat x_n(k|k) &= \hat x_n(k|k - 1) + L_n(k,\gamma,\delta)r_n(k) \hfill
\end{align}
where $r_n(k) := \bar y_a^ - (k) - C\hat x_n(k|k - 1)$; $L_n(k,\gamma,\delta)$ is designed like $L(k,\gamma,\delta)$, \emph{i.e.},
\be
\begin{gathered}
{L_n}(k,\gamma,\delta) = \left( {1 + {\beta _1}\left( {1 - \gamma (k)} \right)} \right){P_{n,k|k - 1}} \hfill \\
\qquad \qquad \qquad \qquad \qquad \qquad \times {C^T}\Psi _{n,\beta }^{ - 1}\left( {k,\gamma,\delta} \right), \hfill \\
\end{gathered}
 \label{eq3A7}
\ee
\be
{P_{n,k|k - 1}} = A{P_{n,k - 1|k - 1}}{A^T} + {\mathcal{E} _w},
\label{eq3A8}
\ee
\begin{align}
  &\label{eq3A9}{\Psi _{n,\beta }}\left( {k,\gamma,\delta} \right) = \left( {1 + {\beta _1}\left( {1 - \gamma (k)} \right)} \right)C{P_{n,k|k - 1}}{C^T} \hfill \\
   &+ \left( {1 + {\beta _2}\left( {1 - \gamma (k)} \right)} \right){\mathcal{E}_v} + \left( {1 - \gamma (k)} \right)\left( {1 + \beta _1^{ - 1} + \beta _2^{ - 1}} \right)\delta \mathds{I}, \hfill \nonumber
\end{align}
\be
{{\bar P}_{n,k|k}} = \left( {1 + {\beta_1}\left( {1 - \gamma (k)} \right)} \right)\left( {\mathds{I} - L_n(k,\gamma,\delta)C} \right){P_{n,k|k - 1}}, \label{eq3A10}
\ee
and ${P_{n,k|k - 1}} := \mathds{E}\left[ {\left. {{e_n}(k|k - 1)e_n^T(k|k - 1)} \right|\gamma (k)} \right]$, ${e_n}(k|k - 1) := x(k) - {{\hat x}_n}(k|k - 1)$; ${P_{n,k|k}} := \mathds{E}\left[ {\left. {{e_n}(k|k)e_n^T(k|k)} \right|\gamma (k)} \right]$ and ${e_n}(k|k) := x(k) - {{\hat x}_n}(k|k)$, ${P_{n,k|k}} \cong {{\bar P}_{n,k|k}}$.

To ensure the stability of the plant, using $\hat x_n(k|k)$ from the above event-triggered estimator, the control signal (\ref{eq217}) deployed by actuator can be re-written as
\be
u(k)=K \hat x_n(k|k).
\label{eq3A11}
\ee

\subsubsection{Design of New Asymptotic ETDW Tests} From the design so far, the new ETDW scheme for ETSE-based NCSs has been completed partially, where there are attack detection formulas remaining to be designed. To design attack detection formulas, the features of $r_n(k)$ and $d_n(k)$ under no attack are presented in the following Theorem~\ref{T3}.
\begin{theorem}
\label{T3}
Considering the system (\ref{eq21})--(\ref{eq24}) and (\ref{eq3A1})--(\ref{eq3A11}) under no attack (\emph{i.e.}, $a(k)=0, \forall k \geqslant 0$), the cross covariance of $r_n(k)$ and $d_n(k)$, and the auto covariance of $r_n(k)$ satisfy
\begin{subequations}
\label{eq3A12}
\begin{align}
\label{eq3A12a}\mathds{E}\left[ {\left. {r_n(k){d_n^T}(k)} \right|\gamma (k)} \right] &= 0, \hfill \\
\label{eq3A12b}\mathds{E}\left[ {\left. {r_n(k){r_n^T}(k)} \right|\gamma (k)} \right] &\leqslant {\Psi _{n,\beta} }\left({k,\gamma,\delta}\right) \hfill
\end{align}
\end{subequations}
where $\Psi_{n,\beta} \left({k,\gamma,\delta}\right)$ is the same as (\ref{eq3A9}).
\end{theorem}

\emph{Proof:} The proof is given in Section \uppercase\expandafter{\romannumeral+3}.C of the supplementary materials.
\endproof

Using (\ref{eq3A12}) in Theorem~\ref{T3}, meanwhile according to the limit convergence theorem in probability \cite[Th. A.6]{GRA}, two new asymptotic ETDW tests can be designed as
\begin{subequations}
\label{eq3A13}
\begin{align}
\label{eq3A13a} \mathop {{\text{p-lim}}}\limits_{i \to \infty } \frac{1}{i}{\mathcal{D}_{n,i}} &= 0, \hfill \\
\label{eq3A13b} \mathop \text{\rm p-lim}\limits_{i \to \infty } \frac{1}{i}{\mathcal{R}_{n,i}} &\leqslant {\Psi _{n,\beta} }\left( {k,\gamma ,\delta} \right) \hfill
\end{align}
\end{subequations}
where the summations are ${\mathcal{D}_{n,i}} := \sum\nolimits_{k = 1}^{i} {r_n(k){d_n^T}(k)}$ and ${\mathcal{R}_{n,i}} := \sum\nolimits_{k = 1}^{i} {r_n(k){r_n^T}(k)}$.

\begin{remark}
The new ETDW tests (\ref{eq3A13}) with the mechanism of watermarking as symmetric key encryption (\ref{eq3A1}) and (\ref{eq3A3}) can detect denial-of-service (DoS) attacks and replay attacks, where the proofs are given in Section~III.C of the supplementary materials. Furthermore, the new ETDW tests against other types of attacks can be investigated in future work.
\end{remark}

\subsubsection{Security Property of New Asymptotic ETDW Tests}
Now, the new ETDW scheme for ETSE-based NCSs is completed. The security property of new ETDW scheme is analysed in the following Theorem~\ref{T4}.
\begin{theorem}
\label{T4}
Considering the system (\ref{eq21})--(\ref{eq24}) and (\ref{eq3A1})--(\ref{eq3A11}), if both (\ref{eq3A13a}) and (\ref{eq3A13b}) are true, then the asymptotic attack power of GRAs is constrained by
\be
\mathop \text{\rm p-lim}\limits_{i \to \infty } \frac{1}{i}\sum\nolimits_{k = 1}^i {{a^T}(k)a(k)} \leqslant tr\left({\Psi _{n,\beta} }\left( {k,\gamma,\delta} \right)\right)
\label{eq3A15}
\ee
where $\Psi_{n,\beta} \left({k,\gamma,\delta}\right)$ is the same as (\ref{eq3A9}).
\end{theorem}

\emph{Proof:} The proof is given in Section \uppercase\expandafter{\romannumeral+3}.D of the supplementary materials.
\endproof
\begin{remark}
Unlike the zero asymptotic power of undetected attacks from trigger-free CDW scheme in Theorem 2, Theorem 3 reveals that with new ETDW scheme used, the asymptotic power of undetected GRAs is not \emph{more than} the power $tr \left({\Psi _{n,\beta} }( {k,\gamma,\delta}) \right)$ of attack-free $r_n(k)$.
\end{remark}

\subsection{Design and Attack Detection Performance Analysis of New Finite Sample ETDW Tests for ETSE-Based NCSs}
We have established a new ETDW scheme for ETSE-based NCSs and analysed its security property. However, the new asymptotic ETDW tests used for new ETDW scheme requires the infinite limit $i \to \infty$ that is unrealistic. To solve the problem, the new finite sample ETDW tests for ETSE-based NCSs are designed and its attack detection performance is analysed as follows.

\subsubsection{New Ideal Finite Sample ETDW Tests}
To construct finite sample ETDW tests, three necessary conditions are required:
\begin{enumerate}
  \item[c1)]
  Make $i$ finite;
  \item[c2)]
  The summation $\mathcal{D}_{n,i}$ and $\mathcal{R}_{n,i}$ need to be formulated;
  \item[c3)]
  To use \emph{matrix concentration inequality} \cite[Prop. 1]{FSDW}, zero-mean matrices need to be defined.
\end{enumerate}
To satisfy the conditions (c1)--(c3), one of the solutions is to derive the expectation of $\mathcal{D}_{n,i}$ or $\mathcal{R}_{n,i}$ from $\mathcal{D}_{n,i}$ or $\mathcal{R}_{n,i}$ respectively, \emph{i.e.},
\begin{align}
  \label{eq3B1}&\frac{1}{i}\left( {{\mathcal{D}_{n,i}} - \sum\nolimits_{k = 1}^i {\mathds{E}\left[ {\left. {{r_n}(k){d_n^T}(k)} \right|\gamma (k)} \right]} } \right), \hfill \\
  \label{eq3B2}&\frac{1}{i}\left( {{\mathcal{R}_{n,i}} - \sum\nolimits_{k = 1}^i {\mathds{E}\left[ {\left. {{r_n}(k)r_n^T(k)} \right|\gamma (k)} \right]} } \right). \hfill
\end{align}
Note that (\ref{eq3B1}) and (\ref{eq3B2}) meet the conditions (c1)--(c3) well. Following Theorem 3, we substitute (\ref{eq3A12a}) and (\ref{eq3A12b}) into (\ref{eq3B1}) and (\ref{eq3B2}) respectively, then,
\begin{align}
  \label{eq3B3}&\frac{1}{i}\left( {{\mathcal{D}_{n,i}} - \sum\nolimits_{k = 1}^i {\mathds{E}\left[ {\left. {{r_n}(k)d_n^T(k)} \right|\gamma (k)} \right]} } \right) = \frac{1}{i}{\mathcal{D}_{n,i}}, \hfill \\
  &\frac{1}{i}\left( {{\mathcal{R}_{n,i}} - \sum\nolimits_{k = 1}^i {\mathds{E}\left[ {\left. {{r_n}(k)r_n^T(k)} \right|\gamma (k)} \right]} } \right) \geqslant \hfill \nonumber\\
  \label{eq3B4}&\qquad \qquad \qquad \quad\frac{1}{i}\left( {{\mathcal{R}_{n,i}} - \sum\nolimits_{k = 1}^i {{\Psi _{n,\beta }}\left( {k,\gamma ,\delta} \right)} } \right). \hfill
\end{align}
It can be seen from (\ref{eq3A12a}) that $\mathds{E}\left[ {\left. {\mathcal{D}_{n,i}} \right|\gamma (k)} \right]=0$ in (\ref{eq3B3}), but the expectation of the right-hand side of (\ref{eq3B4}) is not zero, which is against the condition (c3), \emph{i.e.}, prevents us from using matrix concentration inequality to develop new finite sample ETDW tests.

To cope with the non-zero expectation of the right-hand side of (\ref{eq3B4}), let us firstly focus on the following proposition: If and only if (\ref{eq3A12a}) is true, then $\exists {X_n} \geqslant 0$,
\be
\mathds{E}\left[ {\left. {r_n(k){r_n^T}(k)} \right|\gamma (k)} \right] + X_n ={\Psi _{n,\beta} }\left({k,\gamma,\delta}\right).
\label{eq3B5}
\ee
The proof is omitted.

Using (\ref{eq3B5}), we can define a zero-mean matrix:
\be
\mathcal{\tilde R}_{n,i}^{X_n} := {\mathcal{R}_{n,i}} - \sum\nolimits_{k = 1}^i {{\Psi _{n,\beta} }\left( {k,\gamma,\delta} \right)}  + i X_n
\label{eq3B6}
\ee
where $\mathds{E}\left[ {\left. {\mathcal{\tilde R}_{n,i}^{X_n}} \right|\gamma (k)} \right]=0$. Substituting (\ref{eq3B6}) into the left-hand side of (\ref{eq3B4}) and then using proposition 1, the left-hand side of (\ref{eq3B4}) can be re-written as
\be
\frac{1}{i}\left( {{\mathcal{R}_{n,i}} - \sum\nolimits_{k = 1}^i {\mathds{E}\left[ {\left. {{r_n}(k)r_n^T(k)} \right|\gamma (k)} \right]} } \right)= \frac{1}{i}\mathcal{\tilde R}_{n,i}^{X_n}.
\label{eq3B7}
\ee
For now, (\ref{eq3B3}) and (\ref{eq3B7}) satisfy conditions (c1)--(c3). Furthermore, based on the matrix concentration inequality and using (\ref{eq3B3}) and (\ref{eq3B7}), the new ideal finite sample ETDW tests can be given by the following two events:
\begin{subequations}
\label{eq3B8}
\begin{align}
\label{eq3B8a}{\Xi _{n,1,i}} &:= \left\{\left. {\left\| {\frac{1}{i}{{\mathcal{D}}_{n,i}}} \right\| \geqslant {{\tilde \vartheta }_{n,1,i}}} \right| {\gamma(k)} \right\}, \hfill \\
\label{eq3B8b}{\Xi ^{X_n}_{n,2,i}} &:= \left\{\left. {\left\| {\frac{1}{i} \mathcal{\tilde R}^{X_n}_{n,i}} \right\| \geqslant {{\tilde \vartheta }_{n,2,i}}} \right| {\gamma(k)}\right\} \hfill
\end{align}
\end{subequations}
where detection thresholds ${{\tilde \vartheta }_{n,1,i}} := \sqrt {(1 + {\iota _{n,1}}){\kappa _{n,1}}{{\ln i} \mathord{\left/{\vphantom {{\ln i} i}} \right. \kern-\nulldelimiterspace} i}}$ and ${{\tilde \vartheta }_{n,2,i}} := \sqrt {(1 + {\iota _{n,2}}){\kappa _{n,2}}{{\ln i} \mathord{\left/ {\vphantom {{\ln i} i}} \right. \kern-\nulldelimiterspace} i}}$, and $\iota_{n,1}$, $\iota_{n,2}$, $\kappa_{n,1}$, $\kappa_{n,2}$ are positive scalers.

\subsubsection{New Adding-Threshold Finite Sample ETDW Tests}
Even though the above new ideal finite sample ETDW tests seem to resolve the considered problem directly, there is still a huge gap between (\ref{eq3B8}) and the desired finite sample ETDW tests. This is because $X_n$ used in (\ref{eq3B8b}) is unknown and thus it is impossible to achieve the calculation of (\ref{eq3B8b}) in general. Yet, we observe that the concept of \emph{subset} will help to cope with the considered situation. Specifically, a subset of (\ref{eq3B8b}) allows us to eliminate the matrix ${X_n}$ by adding an adjustable real threshold. Motivated by this, we will construct the new adding-threshold finite sample ETDW tests.

To design the adding-threshold finite sample ETDW tests, the part of $\mathcal{\tilde R}^{X_n}_{n,i}$ in (\ref{eq3B6}) without $X_n$ is defined as
\be
{\mathcal{\tilde R}_{n,i}}: = {\mathcal{R}_{n,i}} - \sum\nolimits_{k = 1}^i {{\Psi _{n,\beta} }} \left( {k,\gamma,\delta} \right)
\label{eq3B9}
\ee
and thus (\ref{eq3B6}) can be re-written as ${\mathcal{\tilde R}^{X_n}_{n,i}} = {\mathcal{\tilde R}_{n,i}}+i X_n$. The relation between $\mathcal{\tilde R}^{X_n}_{n,i}$ and $\mathcal{\tilde R}_{n,i}$ is given in the following Lemma~\ref{T5L1}.
\begin{lemma}
\label{T5L1}
For ${\mathcal{\tilde R}_{n,i}}$ and ${\mathcal{\tilde R}^X_{n,i}}$, there are two cases:
\begin{enumerate}
  \item[i)]
  If $\left\| {\frac{1}{i}{\mathcal{\tilde R}_{n,i}}} \right\| \geqslant {\tilde \vartheta _{n,2,i}} + \Im_n $, then $\left\| {\frac{1}{i}\mathcal{\tilde R}_{n,i}^{X_n}} \right\| \geqslant {{\tilde \vartheta }_{n,2,i}}$;
  \item[ii)]
  If $\left\| {\frac{1}{i}{\mathcal{\tilde R}_{n,i}}} \right\| < {\tilde \vartheta _{n,2,i}} + \Im_n$, it is possible that $\left\| {\frac{1}{i}\mathcal{\tilde R}_{n,i}^{X_n}} \right\| < {{\tilde \vartheta }_{n,2,i}}$ or $\left\| {\frac{1}{i}\mathcal{\tilde R}_{n,i}^{X_n}} \right\| \geqslant {{\tilde \vartheta }_{n,2,i}}$
\end{enumerate}
where $\Im_n = \left\| X_n \right\|$.
\end{lemma}

\emph{Proof:} The proof is given in Section~\uppercase\expandafter{\romannumeral+3}.E in the supplementary materials.
\endproof

According to Lemma~\ref{T5L1}, the adding-threshold finite sample ETDW tests can be given by (\ref{eq3B8a}) and
\be
{\Xi _{n,2,i}} := \left\{\left. {\left\| {\frac{1}{i} \mathcal{\tilde R}_{n,i} } \right\| \geqslant {{\tilde \vartheta }_{n,2,i}}+\tilde \Im_n} \right| {\gamma(k)}\right\}
\label{eq3B10}
\ee
where the added threshold $\tilde \Im_n >0$ is used to approximate $\Im_n$ and needs to be carefully designed, and the detection threshold functions ${{\tilde \vartheta }_{n,1,i}}$ and ${{\tilde \vartheta }_{n,2,i}}$ have been given in (\ref{eq3B8}). Furthermore, we can define
\be
{\Xi _{n,i}} := \left\{{\Xi _{n,1,i}} \cup {\Xi _{n,2,i}}\right\}.
\label{eq3B11}
\ee
We expect that if ${\Xi _{n,i}}$ is true, the attack will be alarmed at $i$-th sampling instant; if $\neg {\Xi _{n,i}}$ is true, there is no attack alarm.

\subsubsection{False Alarm Analysis under no Attack and Detection Performance Analysis for the GRAs of New Adding-Threshold Finite Sample ETDW Tests}
We have fully addressed the new adding-threshold finite sample ETDW tests. The false alarm under no attack of such tests is analysed in the following Theorem~\ref{T5}.(i) based on Lemma~1, and the GRAs detection performance of such tests is presented in the following Theorem~\ref{T5}.(ii).
\begin{theorem}
\label{T5}
Considering the system (\ref{eq21})--(\ref{eq24}) and (\ref{eq3A1})--(\ref{eq3A11}):
\begin{enumerate}
  \item[i)]
  If there is no attack (\emph{i.e.}, $a(k)=0, \forall k \geqslant 0$), $\left\| {w(k)} \right\|<\infty$, $\left\| {v(k)} \right\|<\infty$, and $\tilde \Im_d = \left\| X_n \right\|$, then
  \be
  \mathds{P}\left( {\mathop {\limsup }\limits_{i \to \infty } {\Xi _{n,i}}} \right) = 0
  \label{eq3B112}
  \ee
  \item[ii)]
  If the GRAs do not satisfy (\ref{eq3A15}), $\left\| {w(k)} \right\|<\infty$, $\left\| {v(k)} \right\|<\infty$, and $\tilde \Im_d = \left\| X_n \right\|$, then
  \be
  \mathds{P}\left( {\mathop {\lim \sup }\limits_{i \to \infty } \left. {\neg {\Xi _{n,i}}} \right|s} \right) = 0.
  \label{eq3B13}
  \ee
  \end{enumerate}
\end{theorem}

\emph{Proof:} The proof is given in Section \uppercase\expandafter{\romannumeral+3}.F of the supplementary materials.
\endproof
\begin{remark}
As an event-triggered extension of finite sample CDW tests (\ref{eq219})\cite[Thms. 5 and 7]{FSDW}, Theorem~\ref{T5} reveals that i) under no attack, the new adding-threshold finite time ETDW tests will trigger only finite number of attack alarms for ETSE-based NCSs, and ii) the new adding-threshold finite sample ETDW tests cannot detect the GRAs going against (\ref{eq3A15}) only a finite number of numbers for ETSE-based NCSs, \emph{i.e.}, the GRAs going against (\ref{eq3A15}) can be always detected by such tests.
\end{remark}

\begin{remark}
It is difficult to yield $\tilde \Im_n = \left\| X_n \right\|$ because we do not know the value of $X_n$. But then, by performing multiple attack-free experiments on the system, an appropriate $\tilde \Im_d$ can be selected so that the false alarm under no attack is avoided as much as possible, which is shown in Section~\uppercase\expandafter{\romannumeral+4}.
\end{remark}

\begin{remark}
For now, limitations 1 and 2 of CDW scheme for ETSE-based NCSs have been well-handled as follows:
\[\begin{gathered}
  limitation\ 1:(\ref{eq220})\mathop  \to \limits^{1st} (\ref{eq3A4}), \hfill \\
  limitation\ 2:(\ref{eq221})\mathop  \to \limits^{1st}  (\ref{eq3A12})\mathop  \to \limits^{2nd}  (\ref{eq3A13})\mathop  \to \limits^{2nd}  (\ref{eq3B8})\mathop  \to \limits^{3rd}  (\ref{eq3B8a}), (\ref{eq3B10}). \hfill \\
\end{gathered} \]
Firstly, to overcome limitation 1 (\ref{eq220}) of CDW scheme, the new ETDW scheme guarantees (\ref{eq3A4}) by treating watermarking as symmetric key encryption (\ref{eq3A1}) and (\ref{eq3A3}). Meanwhile, limitation 2 (\ref{eq221}) of CDW scheme is transformed into the feature (\ref{eq3A12}) under new ETDW scheme. Secondly, based on (\ref{eq3A12}), the new asymptotic ETDW tests (\ref{eq3A13}) are designed, which is used to produce the new ideal finite sample ETDW tests (\ref{eq3B8}). Thirdly, based on (\ref{eq3B8}), the new adding-threshold finite sample ETDW tests (\ref{eq3B8a}) and (\ref{eq3B10}) are designed. Finally, Theorems~\ref{T4}, \ref{T5} and experiment results in Section~IV show the reasonable GRAs detection performance of the proposed new ETDW scheme.
\end{remark}

\section{Experiments}
\begin{figure}[!t]
  \centering
  \includegraphics[width=0.4\textwidth]{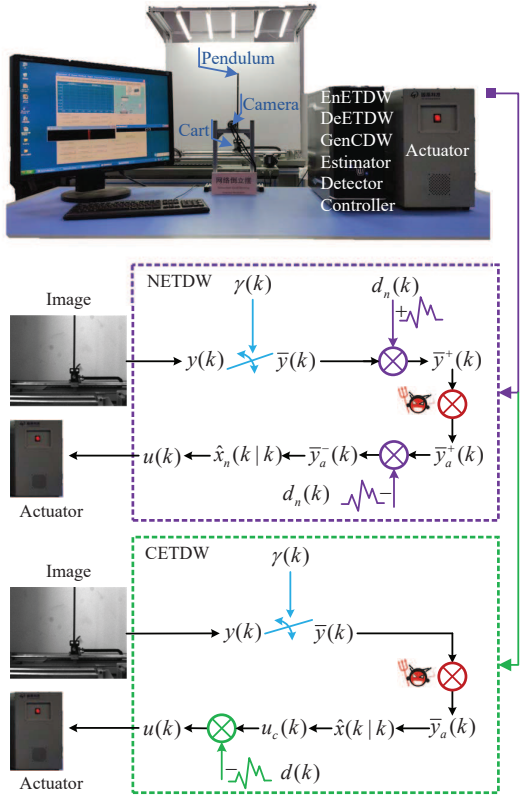}
  \caption{Experimental platform of NIPVSSs with new and candidate ETDW scheme. NETDW: New ETDW scheme. CETDW: Candidate ETDW scheme.}
  \label{fig3}
\end{figure}
To demonstrate the effectiveness of new ETDW scheme, the scenario when the GRAs enter networked inverted pendulum visual servo systems (NIPVSSs) \cite{NIP} with new and candidate ETDW scheme in Section \uppercase\expandafter{\romannumeral+2} of the supplementary materials is considered, as shown in Fig.~\ref{fig3}.

\subsection{Platform of NIPVSS}
The discrete-time system state of NIPVSSs is denoted as $x(k) := \left[ {\alpha (k);\theta (k);\dot \alpha (k);\dot \theta (k)} \right]$, where $\alpha(k)$ and $\theta(k)$ are the cart position and pendulum angle at $k$-th sampling instant, respectively. The state-space model of NIPVSSs is
\[\begin{gathered}
  A = \left[ {\begin{array}{*{20}{c}}
  1&0&{0.0100}&0 \\
  0&{1.0015}&0&{0.0100} \\
  0&0&1&0 \\
  0&{0.2945}&0&{1.0015}
\end{array}} \right],B = \left[ {\begin{array}{*{20}{c}}
  0 \\
  {0.0002} \\
  {0.0100} \\
  {0.0300}
\end{array}} \right] \hfill \\
\end{gathered} \]
and the covariance of $w(k)$ is ${\mathcal{E} _w} = diag\left\{ {0,0,{{10}^{ - 5}},{{10}^{ - 5}}} \right\}$. The system output is $y(k)=\left[{\alpha(k);\theta(k)}\right]$, thus $C = \left[ {\mathds{I},0} \right]$; the covariance of $v(k)$ is ${\mathcal{E} _v} = diag\left\{ {2.7 \times {{10}^{ - 7}},5.5 \times {{10}^{ - 6}}} \right\}$.

The trigger (\ref{eq23}) is set as $\delta  = 0.00001$. The event-triggered estimators (\ref{eq210}), (\ref{eq211}), (\ref{eq3A5}) and (\ref{eq3A6}) are set as ${P_{ - 1| - 1}} = {P_{n, - 1| - 1}}= 0$, ${\beta _1} = {\beta _2} = 0.02$. By selecting $Q=diag\{10,10,10,10\}$ and $R=1$, the controllers (\ref{eq217}) and (\ref{eq3A11}) are designed as $K=[2.8889,-36.6415,4.9141,-7.3267]$. To analyse the power of the GRAs, a quantity of attack power is defined as $\mathcal{A}(i) = \frac{1}{i}\sum\nolimits_{k = 1}^i {{a^T}(k)a(k)}$.

Considering the limits of reality, there are two bounds: $\left| {\alpha (k)} \right| \leqslant 0.3m$, $\left| {\theta (k)} \right| \leqslant 0.8rad$. Once one of the above two bounds is crossed, the servo motor of NIPVSSs will be put ``OFF'', \emph{i.e.}, NIPVSSs will get out of control.

\subsection{CDW Scheme on TTC}
To analyse the experimental results of NIPVSSs with CDW scheme on TTC, the following two steps are performed.

Step 1: Six experiments on NIPVSSs with CDW scheme of $\mathcal{E}_{d}=0.01$ on TTC under no attack are carried out. To save page, one of six experiments is shown in Fig.~A.1 of Section~\uppercase\expandafter{\romannumeral+4}.A in the supplementary materials. Six experiments are used to determine the detection threshold functions for CDW tests on TTC with $\mathcal{E}_{d}=0.01$. The results of attack-free CDW tests on TTC with $\mathcal{E}_{d}=0.01$ are shown in Fig.~A.2, and the value of $\mathcal{E}_r^{f}$ and the concrete parameters of $\vartheta_{1,i}$ and $\vartheta_{2,i}$ are given in Section \uppercase\expandafter{\romannumeral+4}.A of the supplementary materials.

\begin{figure}[!t]
  \centering
  \subfigure[Cart position]{\includegraphics[width=0.4\textwidth]{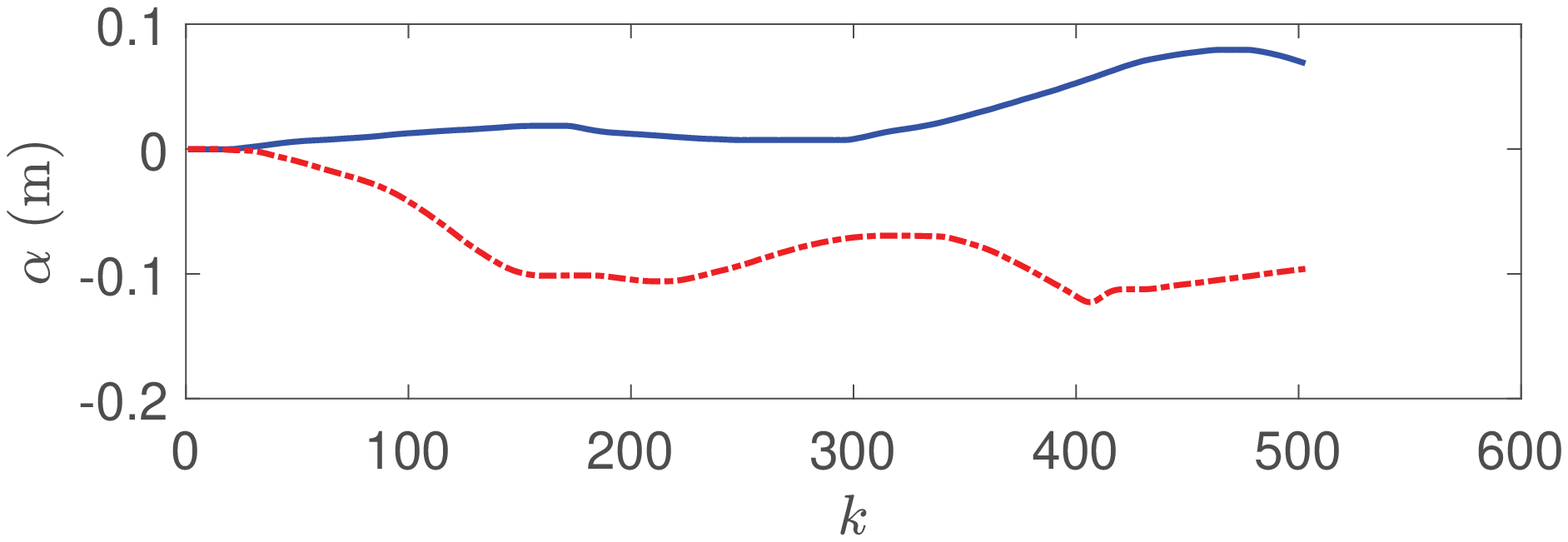}}  \\ \vspace{-0.15in}
  \subfigure[Pendulum angle]{\includegraphics[width=0.4\textwidth]{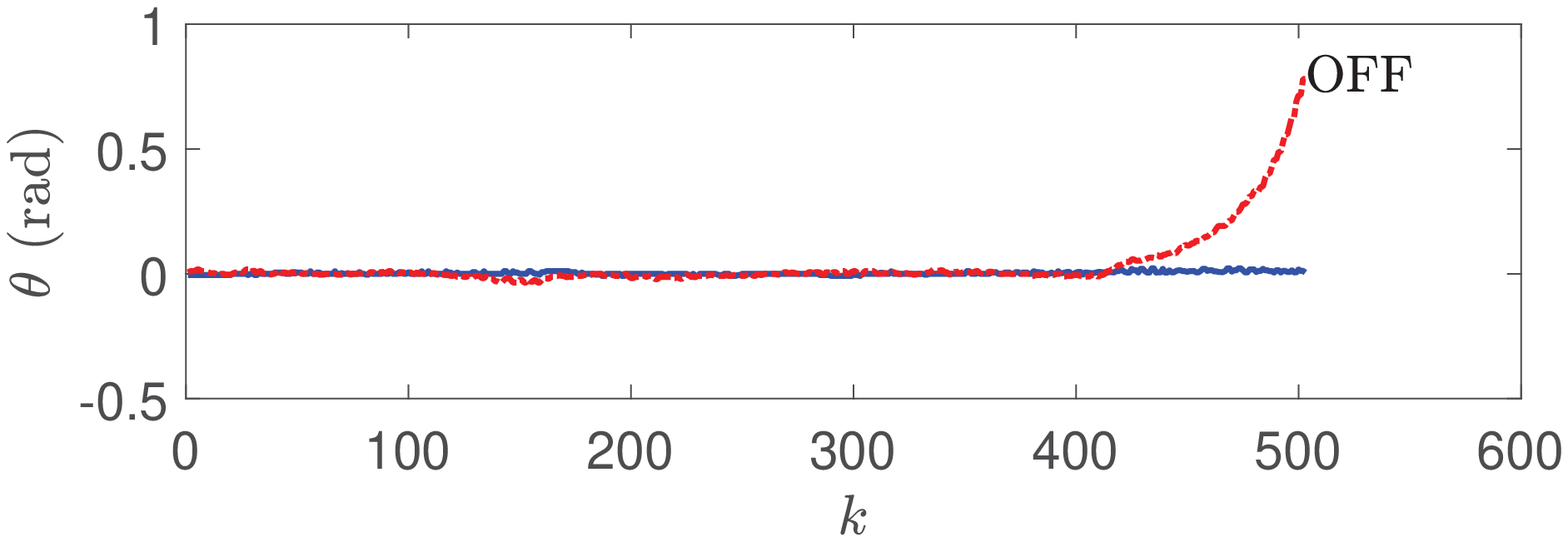}} \\ \vspace{-0.15in}
  \subfigure[$\mathcal{A}(k)$]{\includegraphics[width=0.4\textwidth]{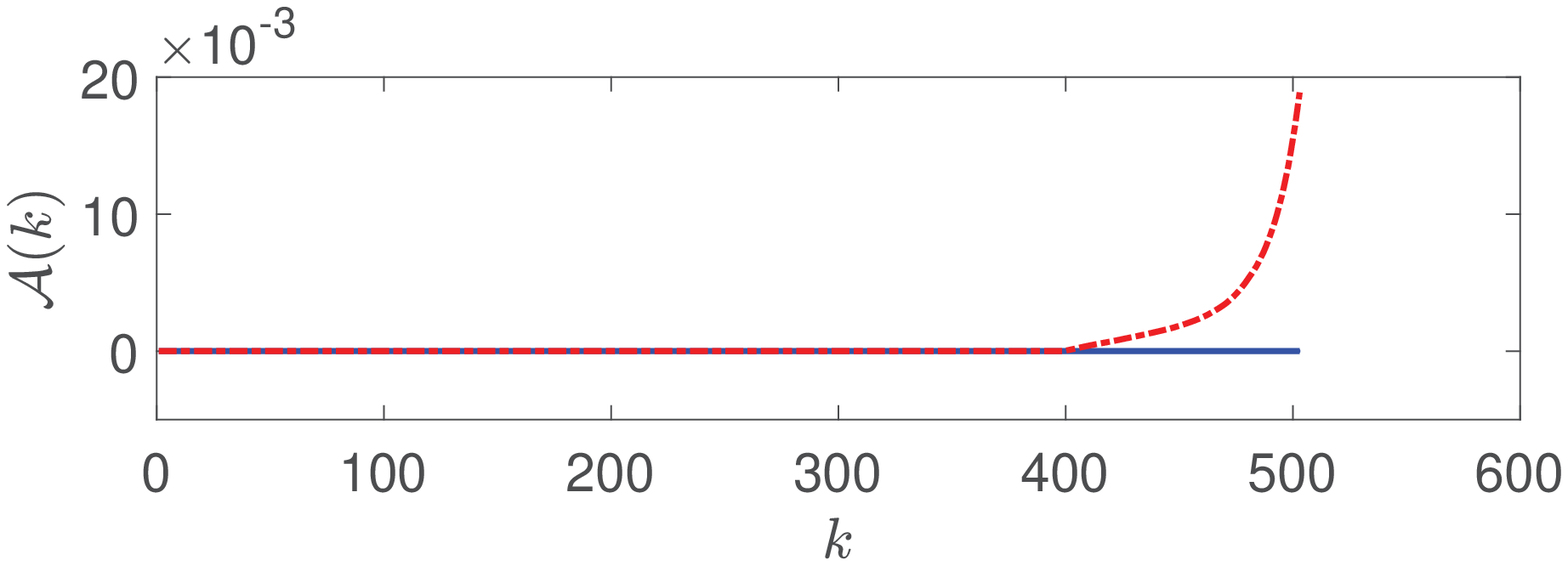}} \\ \vspace{-0.15in}
  \subfigure[$\gamma(k)$]{\includegraphics[width=0.4\textwidth]{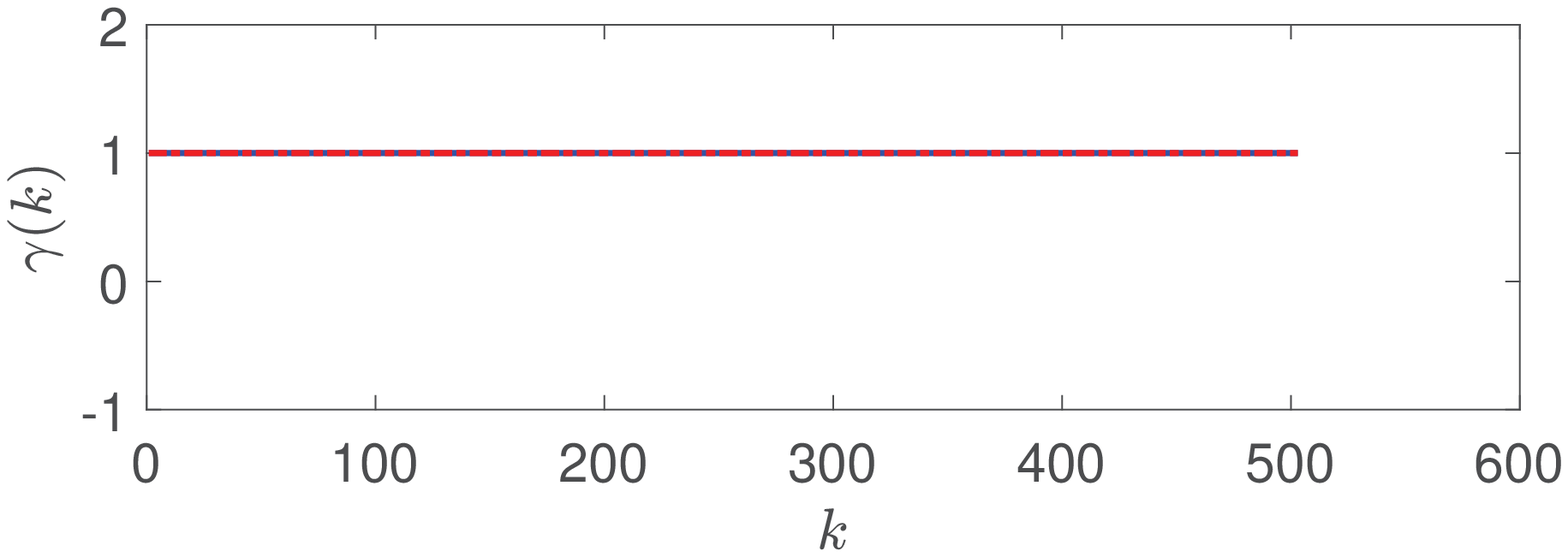}} \\ \vspace{-0.15in}
  \subfigure[$\frac{1}{i}{\mathcal{D}_i}$]{\includegraphics[width=0.4\textwidth]{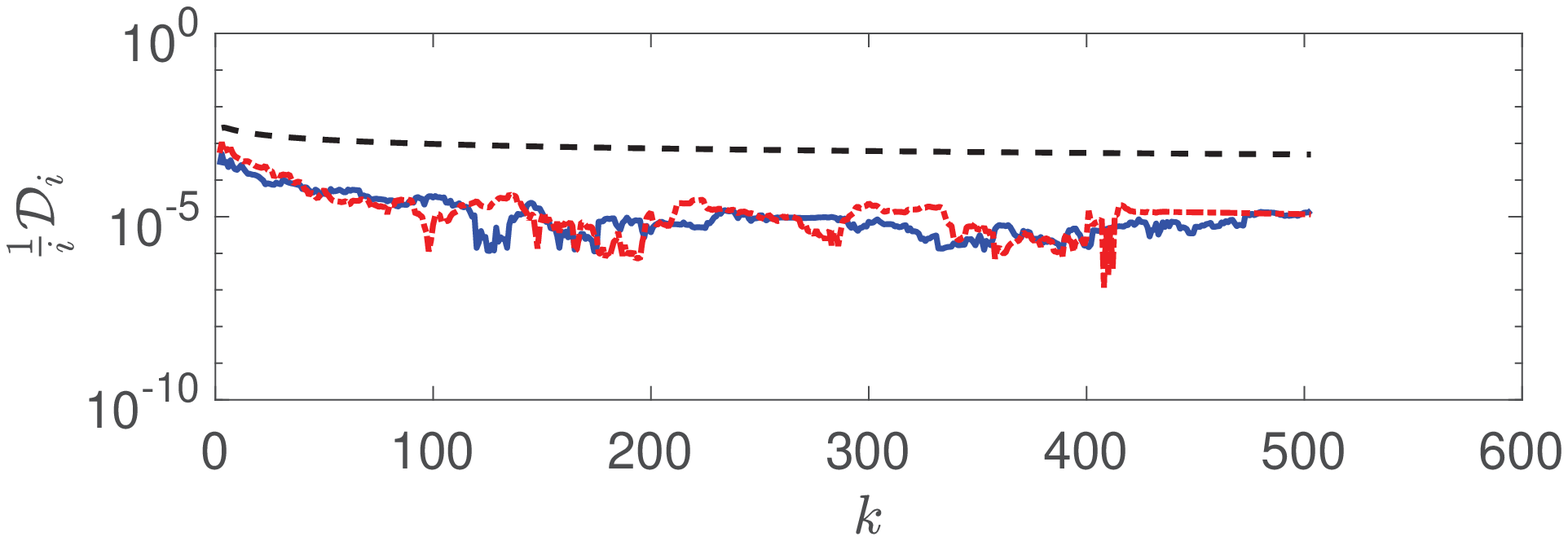}} \\ \vspace{-0.15in}
  \subfigure[$\frac{1}{i}{(\mathcal{R}_i-i\mathcal{E}_r^f)}$]{\includegraphics[width=0.4\textwidth]{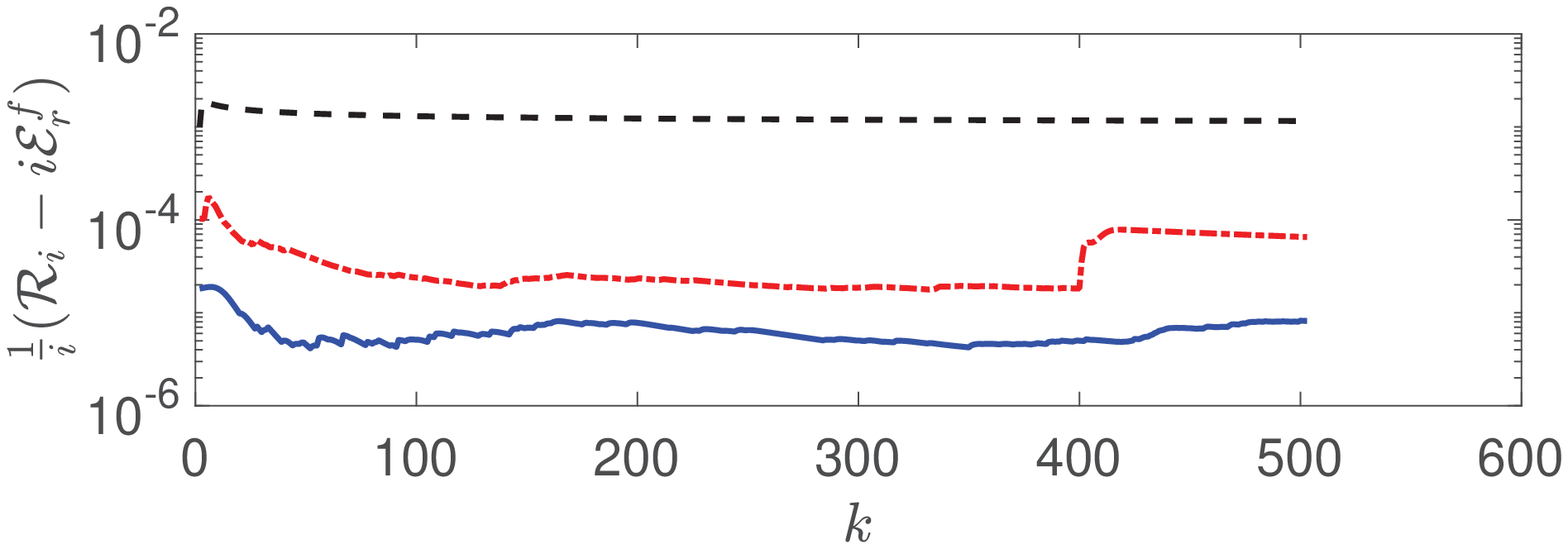}} \\
  \caption{States, attack power, triggering signal and detection results of NIPVSSs with CDW scheme on TTC of $\mathcal{E}_d=0.01$ under the GRAs from $k \geqslant 400$. (a), (b), (d)-(f): Blue Line, attack-free; Red Line, under the GRAs; Black Line, detection threshold function. (c): Blue Line, zero line; Red Line, attack power $\mathcal{A}(k)$.}
  \label{fig4}
\end{figure}
Step 2: we construct the GRAs with $s=-1$, $v_a(k)=0$ and $A_a=diag\left\{{0.1,0.1,0.1,0.1}\right\}$ from $k \geqslant 400$. The detection results of NIPVSSs with CDW scheme on TTC of $\mathcal{E}_d=0.01$ under the GRAs are shown in Fig.~\ref{fig4}, where it can be seen that 1) the pendulum angle is driven to cross 0.8 rad, and the corresponding attack power $\mathcal{A}(k)$ considerably exceeds zero base line, and 2) the CDW tests on TTC of $\mathcal{E}_d=0.01$ fail to detect the GRAs.

\subsection{Triggering Rates of ETC}
\begin{table}[!t]
\centering
\caption{Triggering Rates Analysis of Attack-Free NIPVSSs with New and Candidate ETDW Scheme when $\delta=0.00001$}
\label{Tab2}
\begin{tabular}{lccc}
\toprule
  ~            & TR for NETDW$^1$ & ~ &TR for CETDW$^2$ \\
\midrule
  1st Exp.$^3$ & 38.6008\% & 7th Exp. & 44.1856\%  \\
  2nd Exp.     & 43.6083\% & 8th Exp. & 40.4984\%  \\
  3rd Exp.     & 39.9198\% & 9th Exp. & 40.5813\%  \\
  4th Exp.     & 40.7949\% & 10th Exp. & 38.5386\%  \\
  5th Exp.     & 42.5484\% & 11th Exp. & 45.3508\%  \\
  6th Exp.     & 41.1089\% & 12th Exp. & 37.4888\%  \\
\bottomrule
\multicolumn{4}{l}{$^1$New ETDW. $^2$Candidate ETDW. $^3$Experiment.}\\
\end{tabular}
\end{table}
To analyse the triggering rates (TRs) of ETC with $\delta=0.00001$, we perform twelve experiments on NIPVSSs with new ETDW scheme ($\mathcal{E}_{d_n}=0.01\mathds{I}$, six times) and candidate ETDW scheme ($\mathcal{E}_d=0.01$, six times) under no attack. To save page, two of twelve experiments are shown in Figs.~A.3 and~A.4 of Section~\uppercase\expandafter{\romannumeral+4}.B in the supplementary materials, and the values of TRs are shown in Table~\ref{Tab2}, where 1) it can be calculated that when $\delta=0.00001$, the average TRs of NIPVSSs with new ETDW scheme of $\mathcal{E}_{d_n}=0.01\mathds{I}$ (six times) and candidate ETDW scheme of $\mathcal{E}_d=0.01$ (six times) are 41.0969\% and 41.1073\% respectively and it means that the trigger can save much communication resource, and 2) the trigger can make NIPVSSs operate stably.

\subsection{Detection Threshold Functions and False Alarm for Candidate and New ETDW Tests on ETC}
\begin{table}[!t]
\centering
\caption{Concrete Parameters in $\tilde \vartheta_{1,n,i}$ and $\tilde \vartheta_{2,n,i}+\tilde \Im_n$ ($\tilde \vartheta_{1,i}$ and $\tilde \vartheta_{2,i} + \tilde \Im$)}
\label{Tabthd}
\begin{tabular}{cccccc}
\toprule
   ~&\multicolumn{2}{c}{\tabincell{c}{$\tilde \vartheta_{1,n,i}$\\($\tilde \vartheta_{1,i}$)}} & \multicolumn{3}{c}{\tabincell{c}{$\tilde \vartheta_{2,n,i}+\tilde \Im_n$\\($\tilde \vartheta_{2,i} + \tilde \Im$)}}\\
   \cmidrule(r){2-3} \cmidrule(r){4-6}
   ~&\tabincell{c}{$\iota_{n,1}$\\($\iota_{1}$)} & \tabincell{c}{$\kappa_{n,1}$\\($\kappa_{1}$)} &\tabincell{c}{$\iota_{n,2}$\\($\iota_{2}$)} & \tabincell{c}{$\kappa_{n,2}$($\kappa_{2}$)} & \tabincell{c}{$\tilde \Im_n$($\tilde \Im$)}\\
\midrule
    NETDW$^1$      & 1.0   & 1.8e-7   & 1.0   & 1.0e-6   & 1.0e-3\\
   (CETDW$^2$)     & (1.0) & (1.0e-5) & (1.0) & (1.0e-6) & (1.0e-3)\\
\bottomrule
\multicolumn{6}{l}{$^1$Candidate ETDW. $^2$New ETDW.}\\
\end{tabular}
\end{table}
The above 12 experiments are also used to determine the detection threshold functions for new and candidate finite sample adding-threshold ETDW tests with any $\mathcal{E}_{d_n}$ and $\mathcal{E}_d = 0.01$ respectively. The results of candidate and new finite sample adding-threshold tests are shown in Figs.~A.5-A.7 of Section \uppercase\expandafter{\romannumeral+4}.C in the supplementary materials, and the concrete parameters of detection threshold functions in $\tilde \vartheta_{1,i}$, $\tilde \vartheta_{2,i} + \tilde \Im$, $\tilde \vartheta_{1,n,i}$ and $\tilde \vartheta_{2,n,i} + \tilde \Im_d$ are given in Table~\ref{Tabthd}, where it can be clearly seen that the false alarm under no attacks can be avoided as much as possible by selecting appropriate $\tilde \vartheta_{1,i}$, $\tilde \vartheta_{2,i} + \tilde \Im$, $\tilde \vartheta_{1,n,i}$ and $\tilde \vartheta_{2,n,i} + \tilde \Im_n$.

\subsection{The GRAs Detection Effectiveness for Candidate and New ETDW Tests on ETC}
\begin{figure}[!t]
  \centering
  \subfigure[Cart position]{\includegraphics[width=0.4\textwidth]{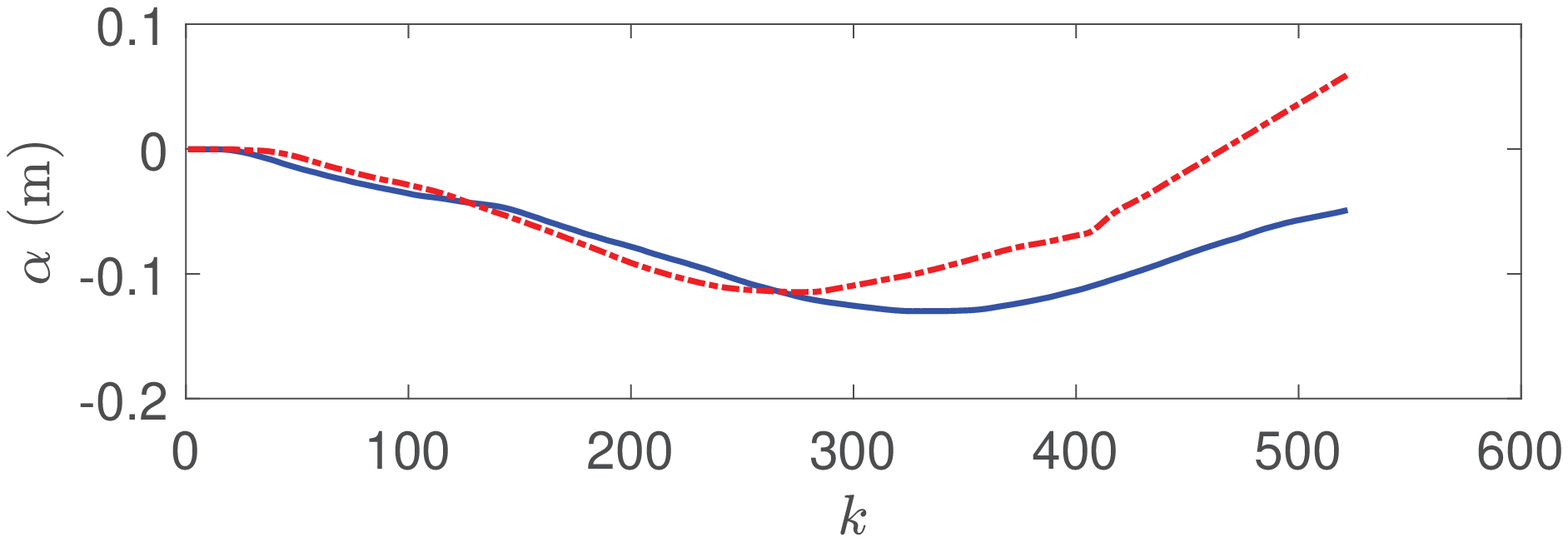}}  \\ \vspace{-0.15in}
  \subfigure[Pendulum angle]{\includegraphics[width=0.4\textwidth]{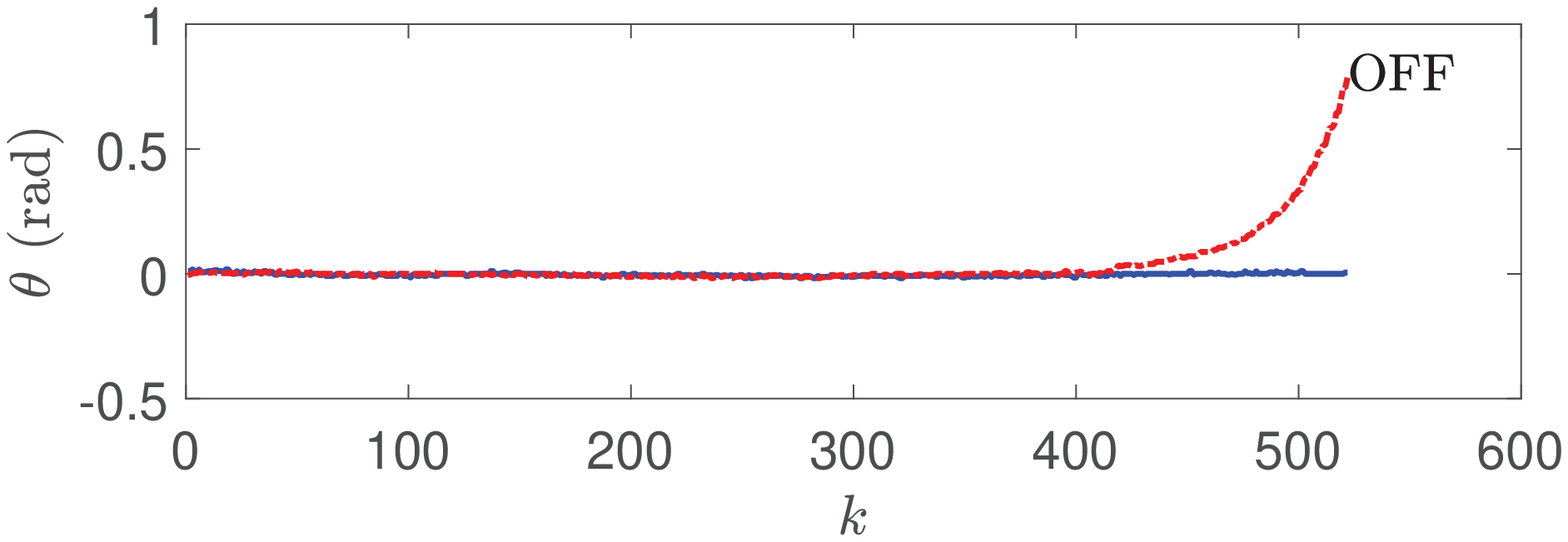}} \\ \vspace{-0.15in}
    \subfigure[$\mathcal{A}(k)$]{\includegraphics[width=0.4\textwidth]{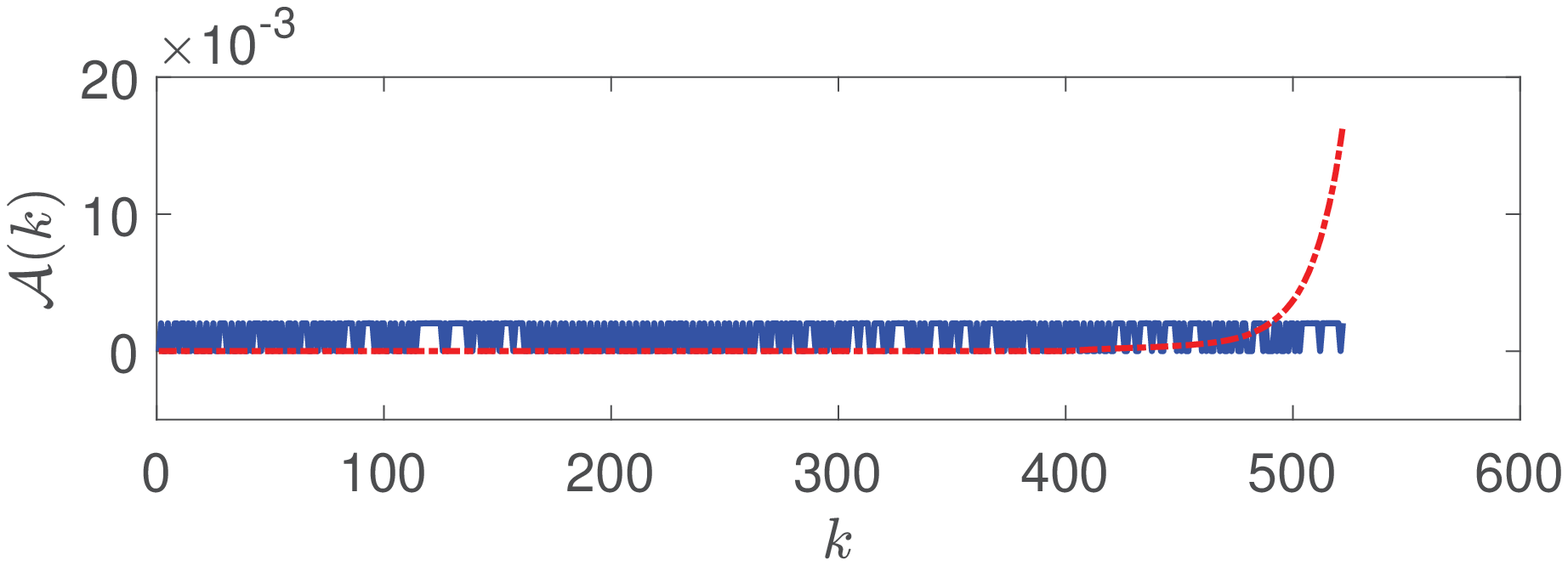}} \\ \vspace{-0.15in}
  \subfigure[$\gamma(k)$]{\includegraphics[width=0.4\textwidth]{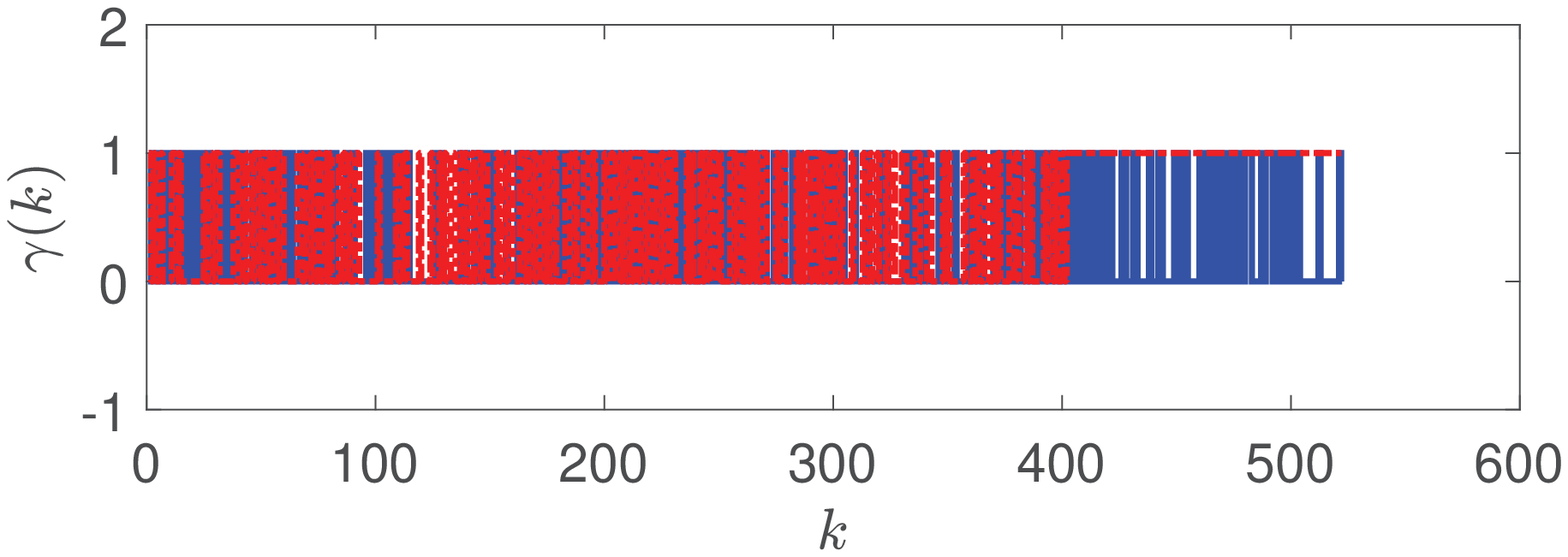}} \\ \vspace{-0.15in}
  \subfigure[$\frac{1}{i}{\mathcal{D}_i}$]{\includegraphics[width=0.4\textwidth]{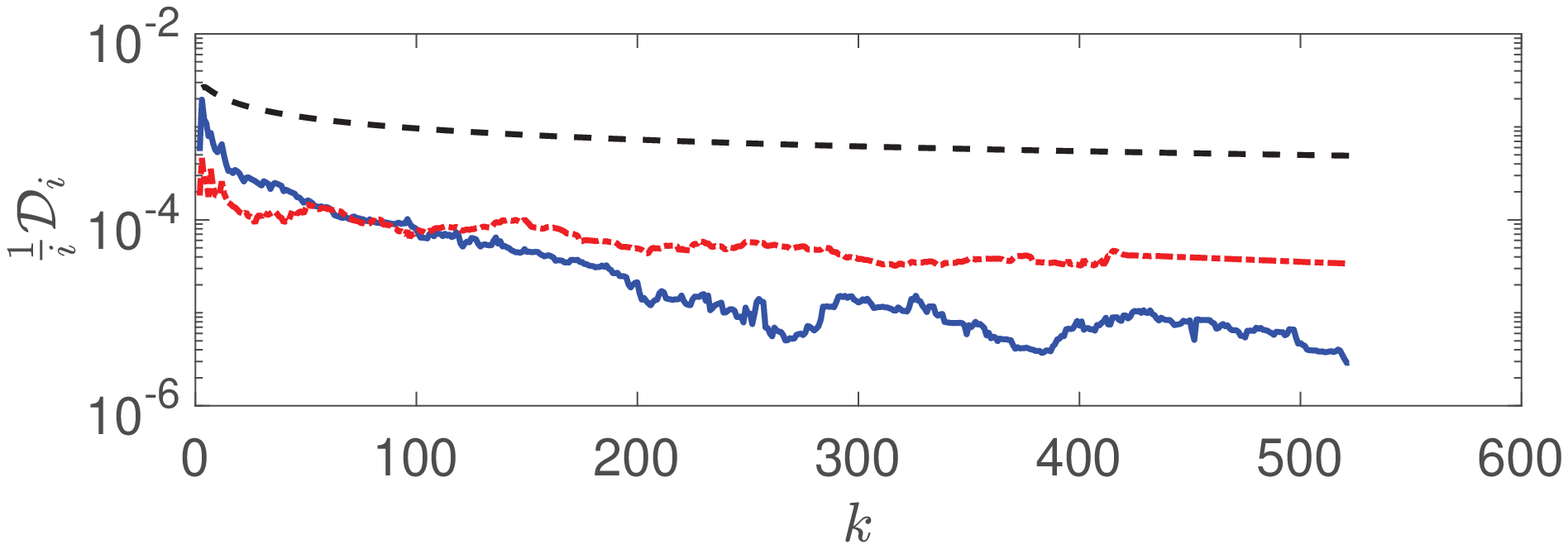}} \\ \vspace{-0.15in}
  \subfigure[$\frac{1}{i}{\mathcal{\tilde R}_i}$]{\includegraphics[width=0.4\textwidth]{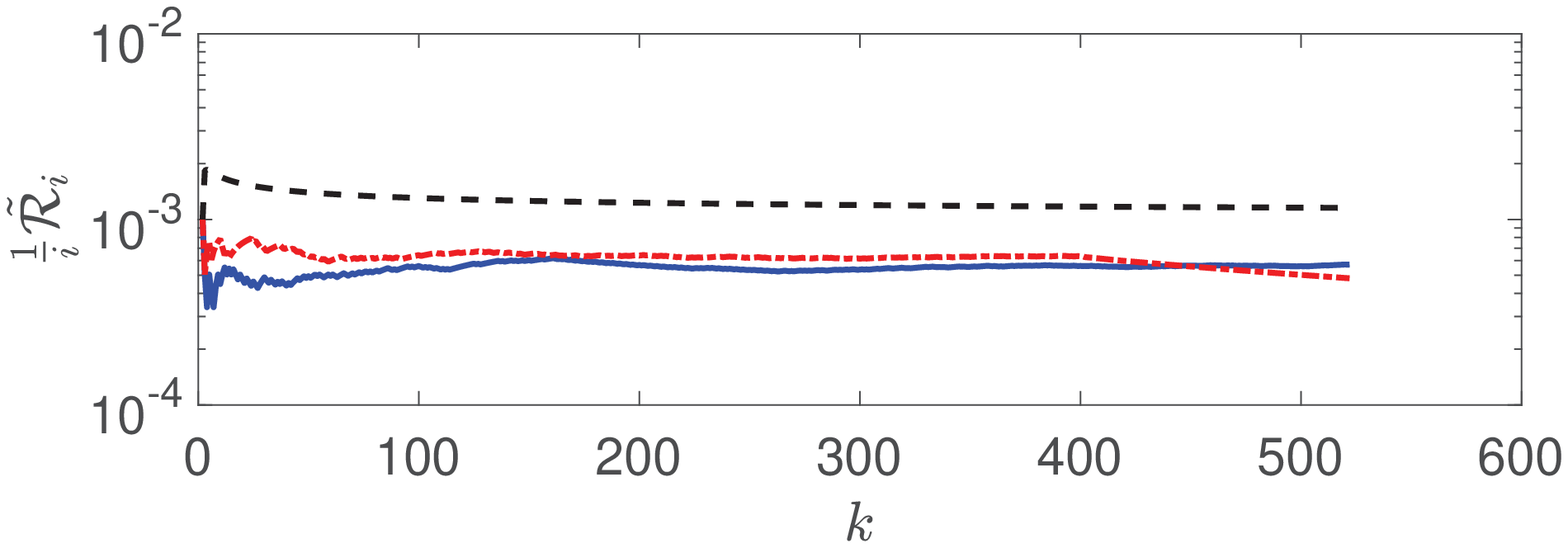}} \\
  \caption{States, attack power, triggering signal and detection results of NIPVSSs with candidate ETDW scheme of $\mathcal{E}_d=0.01$ under the GRAs from $k \geqslant 400$. (a), (b), (d)-(f): Blue Line, attack-free; Red Line, under the GRAs; Black Line, detection threshold function. (c): Blue Line, $tr\left( {{\Psi _\beta }(k,\gamma,\delta)} \right)$; Red Line, attack power $\mathcal{A}(k)$.}
  \label{fig5}
\end{figure}
\begin{figure}[!t]
  \centering
  \subfigure[Cart position]{\includegraphics[width=0.4\textwidth]{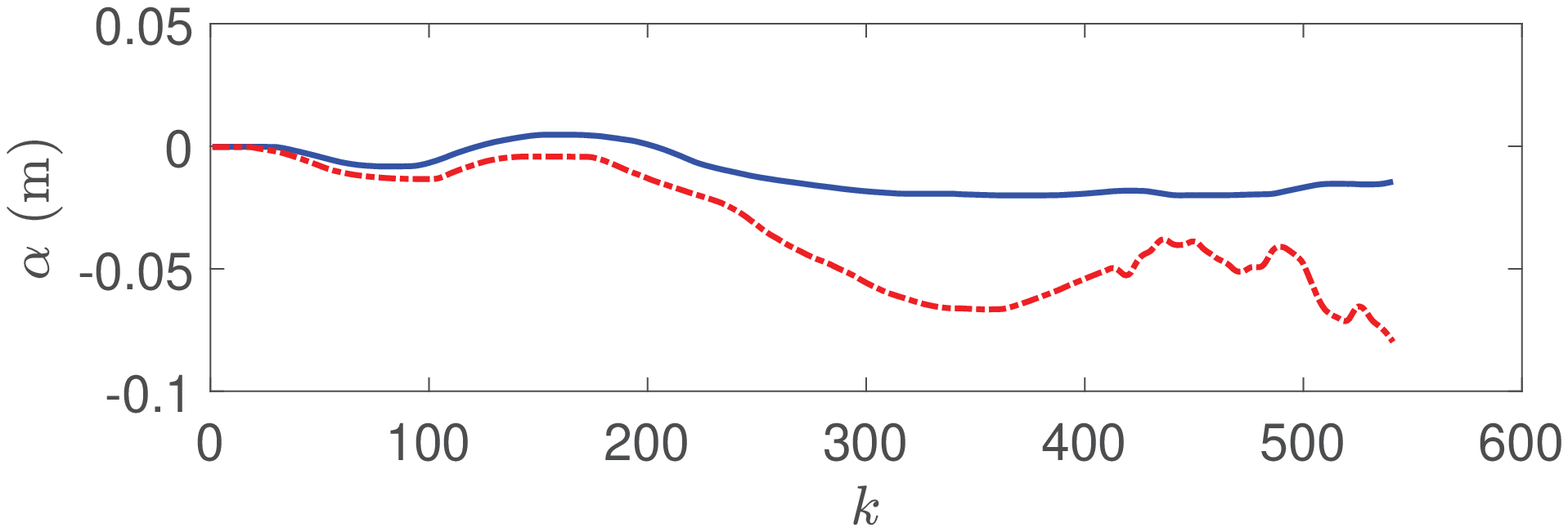}}  \\ \vspace{-0.15in}
  \subfigure[Pendulum angle]{\includegraphics[width=0.4\textwidth]{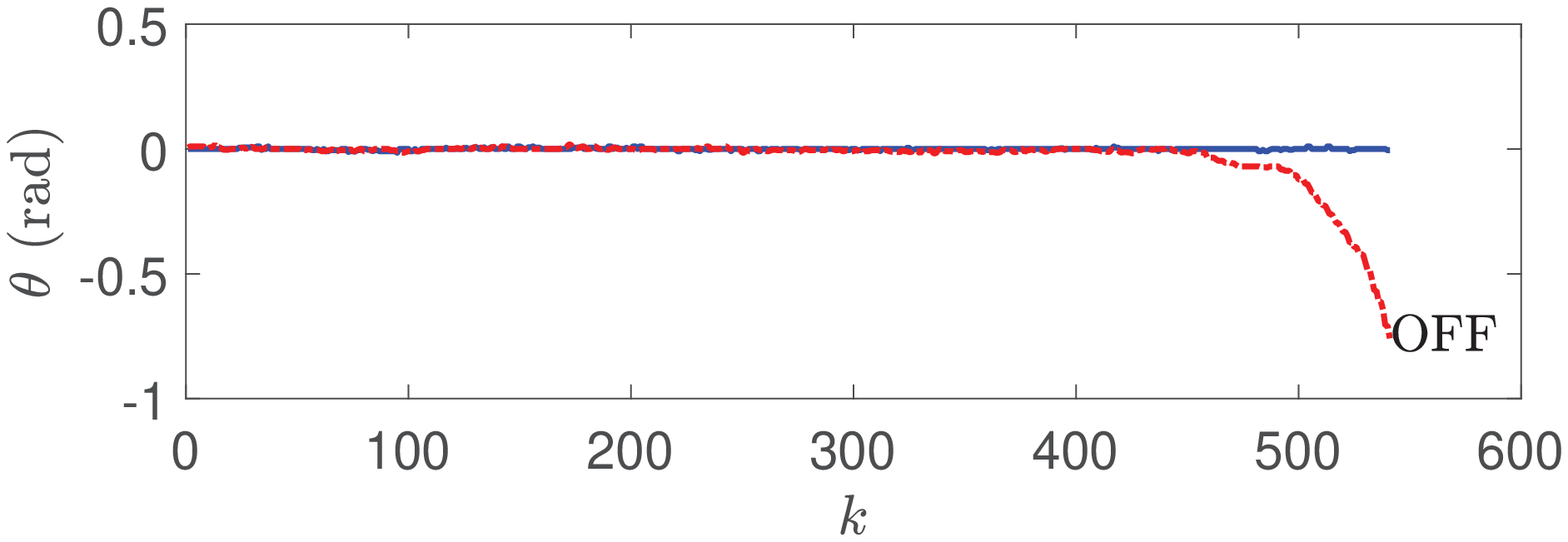}} \\ \vspace{-0.15in}
  \subfigure[$\mathcal{A}(k)$]{\includegraphics[width=0.4\textwidth]{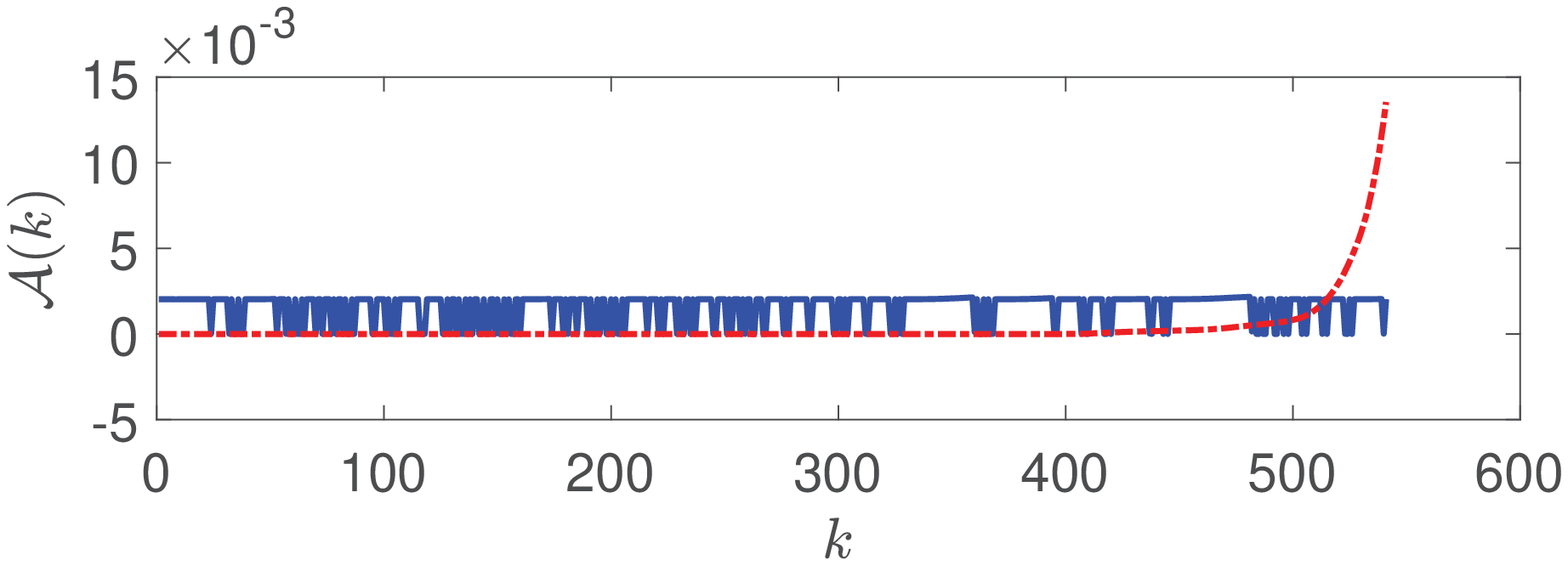}} \\ \vspace{-0.15in}
  \subfigure[$\gamma(k)$]{\includegraphics[width=0.4\textwidth]{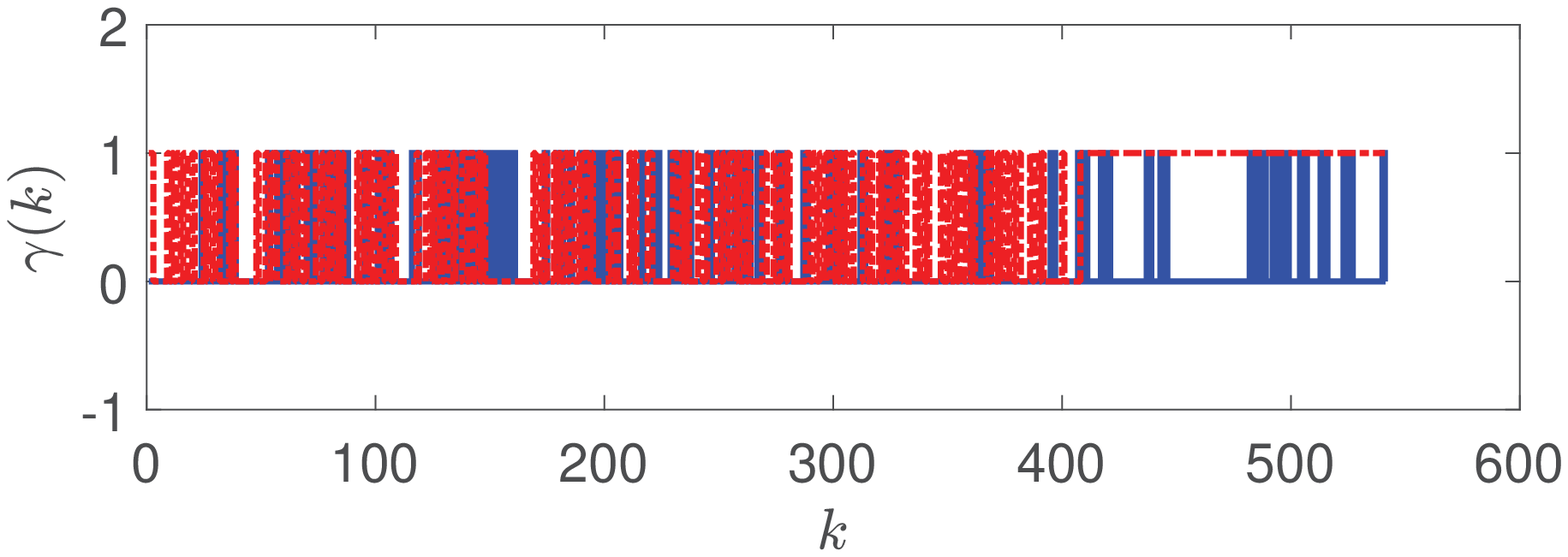}} \\ \vspace{-0.15in}
  \subfigure[$\frac{1}{i}{\mathcal{D}_{n,i}}$]{\includegraphics[width=0.4\textwidth]{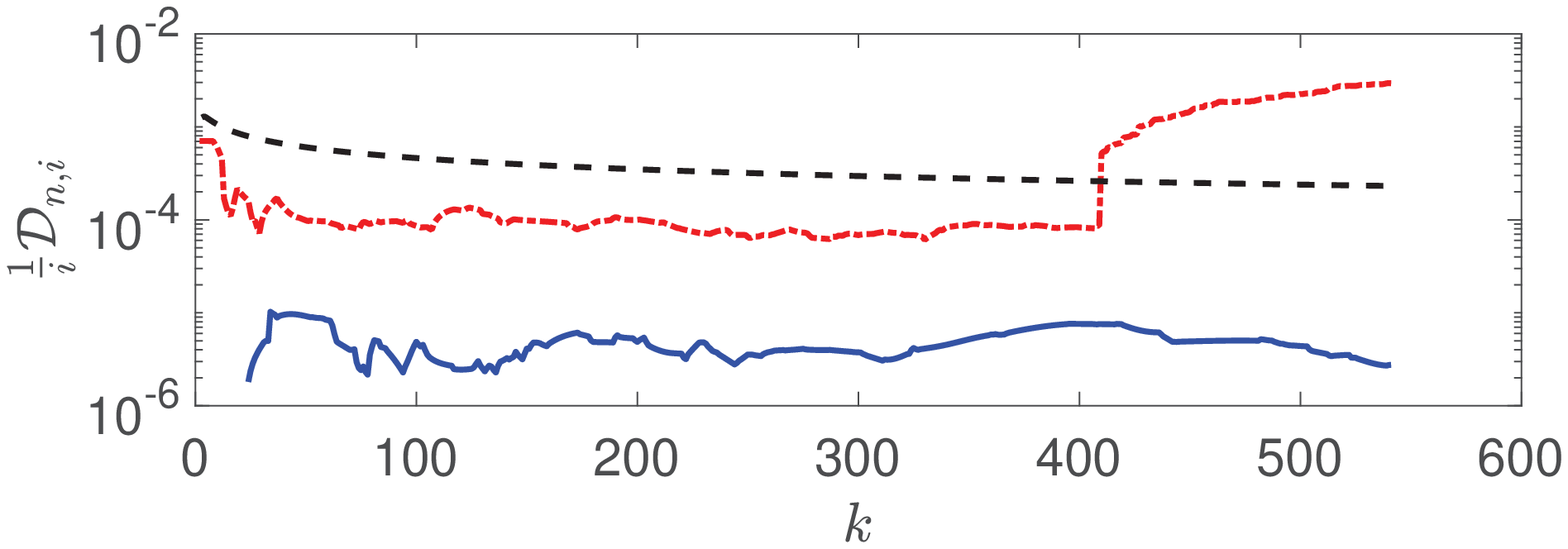}} \\ \vspace{-0.15in}
  \subfigure[$\frac{1}{i}{\mathcal{\tilde R}_{n,i}}$]{\includegraphics[width=0.4\textwidth]{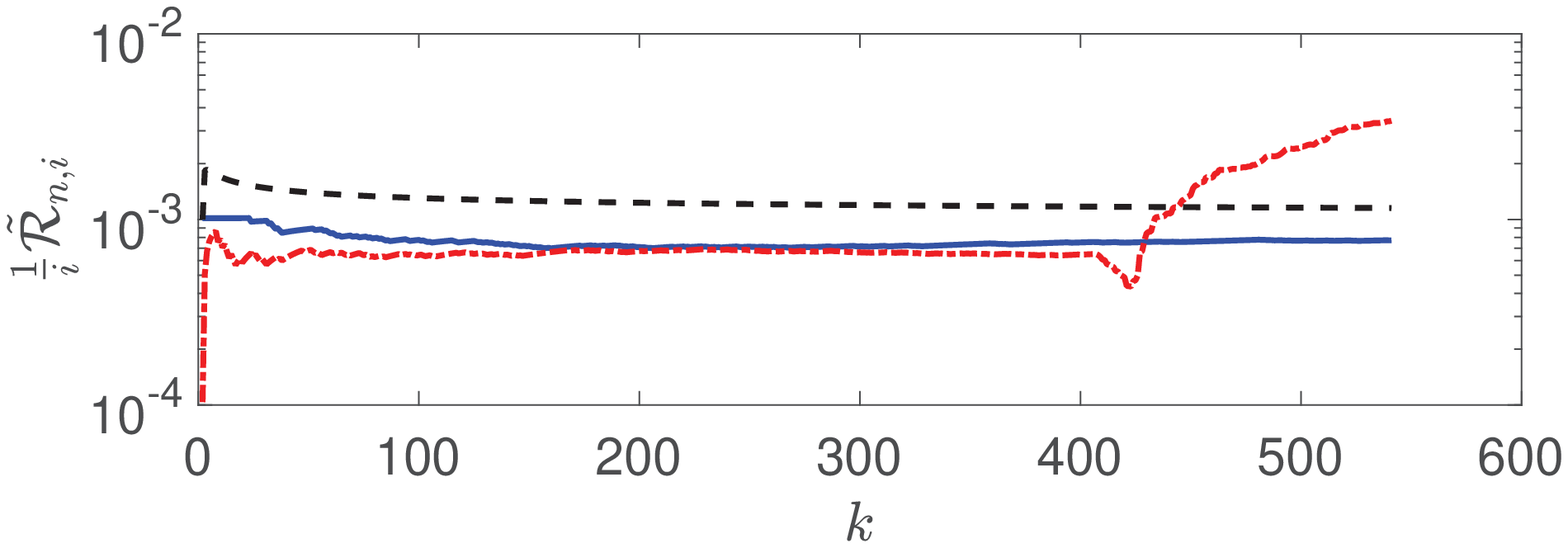}} \\
  \caption{States, attack power, triggering signal and detection results of NIPVSSs with new ETDW scheme of $\mathcal{E}_{d_n}=0.01\mathds{I}$ under the GRAs from $k \geqslant 400$. (a), (b), (d)-(f): Blue Line, attack-free; Red Line, under the GRAs; Black Line, detection threshold function. (c): Blue Line, $tr\left( {{\Psi _{n,\beta }}(k,\gamma,\delta)} \right)$; Red Line, attack power $\mathcal{A}(k)$.}
  \label{fig6}
\end{figure}
We construct the GRAs with $s=-1$, $v_a(k)=0$ and $A_a=diag\left\{{0.1,0.1,0.1,0.1}\right\}$ from $k \geqslant 400$. The detection results of NIPVSSs with candidate finite sample adding-threshold ETDW tests of $\mathcal{E}_d=0.01$ and new finite sample adding-threshold ETDW tests of $\mathcal{E}_{d_n}=0.01\mathds{I}$ under the GRAs are shown in Figs.~\ref{fig5} and \ref{fig6} respectively, where it can be seen that 1) the candidate finite sample adding-threshold ETDW tests of $\mathcal{E}_d=0.01$ fails to detect the GRAs and the pendulum angle is driven to cross 0.8 rad, and the corresponding attack power $\mathcal{A}(k)$ considerably exceeds the value of $tr\left( {{\Psi _\beta }(k,\gamma,\delta)} \right)$; and 2) new finite sample adding-threshold ETDW tests of $\mathcal{E}_{d_n}=0.01\mathds{I}$ succeed to detect the GRAs and the pendulum angle is driven to cross 0.8 rad, and the corresponding attack power $\mathcal{A}(k)$ considerably exceeds the value of $tr\left( {{\Psi _{n,\beta} }(k,\gamma,\delta)} \right)$.

\begin{remark}
It can be clearly seen from Section IV.B-E that the time-triggered strategy provides worse detection performance than the event-triggered one. Specifically, the CDW tests on TTC of ${\mathcal{E}_d} = 0.01$ fail to report the GRAs as shown in Fig.~\ref{fig4}, while the ETDW tests of ${\mathcal{E}_{{d_n}}} = 0.01\mathds{I}$ succeed to report the GRAs as shown in Fig.~\ref{fig6}. With the bigger ${\mathcal{E}_d}$ (\emph{e.g.}, ${\mathcal{E}_d} = 10$), the CDW tests on TTC could succeed to report the GRAs. However, large ${\mathcal{E}_d}$ for the CDW tests will degrade the control performance, or even make systems crash. In comparison with the CDW tests, the ETDW tests with watermarking signals of arbitrary ${\mathcal{E}_{{d_n}}}$ will not degrade the control performance as shown in (\ref{eq3A4}), while the bigger $\delta$ (\emph{i.e.}, the lower triggering frequency) will decline the control performance. Therefore, co-design between the event-triggered threshold $\delta$ and the controller gain $K$ provides a path to guarantee the control performance.
\end{remark}

\subsection{The GRAs Detection Effectiveness for New ETDW Tests from Watermarking Intensity on ETC}
To further investigate the impact of $\mathcal{E}_{d_n}$ on the GRAs detection effectiveness, we perform the same experiments of Fig.~\ref{fig6} again with new adding-threshold finite sample ETDW tests of $\mathcal{E}_{d_n}=0.0001\mathds{I}$. The experiment results are shown in Figs.~A.8 of Section \uppercase\expandafter{\romannumeral+4}.D in the supplementary materials, where it can be seen that new adding-threshold finite sample ETDW tests of $\mathcal{E}_{d_n}=0.0001\mathds{I}$ fail to detect the GRAs (in which the pendulum angle is driven to cross 0.8 rad). Therefore, a big enough watermarking intensity (\emph{e.g.}, $\mathcal{E}_{d_n}=0.01\mathds{I}$) should be selected in new adding-threshold finite sample ETDW tests for successful attack detection.

\section{Conclusion}
A linear event-triggered extension to the CDW scheme had been developed. Specifically, a new ETDW scheme was designed from new asymptotic ETDW tests to new ideal finite sample ETDW tests, which limited the power of undetected GRAs and guaranteed the finite false alarm under no attacks and finite failures on GRAs detection. Experimental results on NIPVSSs verified the proposed scheme.

The ETDW scheme can be used in nonlinear systems. According to the approach of linearization, the ETDW scheme can be directly applied to the linearized nonlinear system with low degree of nonlinearity. However, for nonlinear systems with high degree of nonlinearity, it has to incorporate the properties of saturation, dead zone, gap, relay and so forth to analyze the covariance of signals. Therefore, it is interesting to extend ETDW scheme into nonlinear systems in future work.




\begin{IEEEbiography}[{\includegraphics[width=1in,height=1.25in,clip,keepaspectratio]{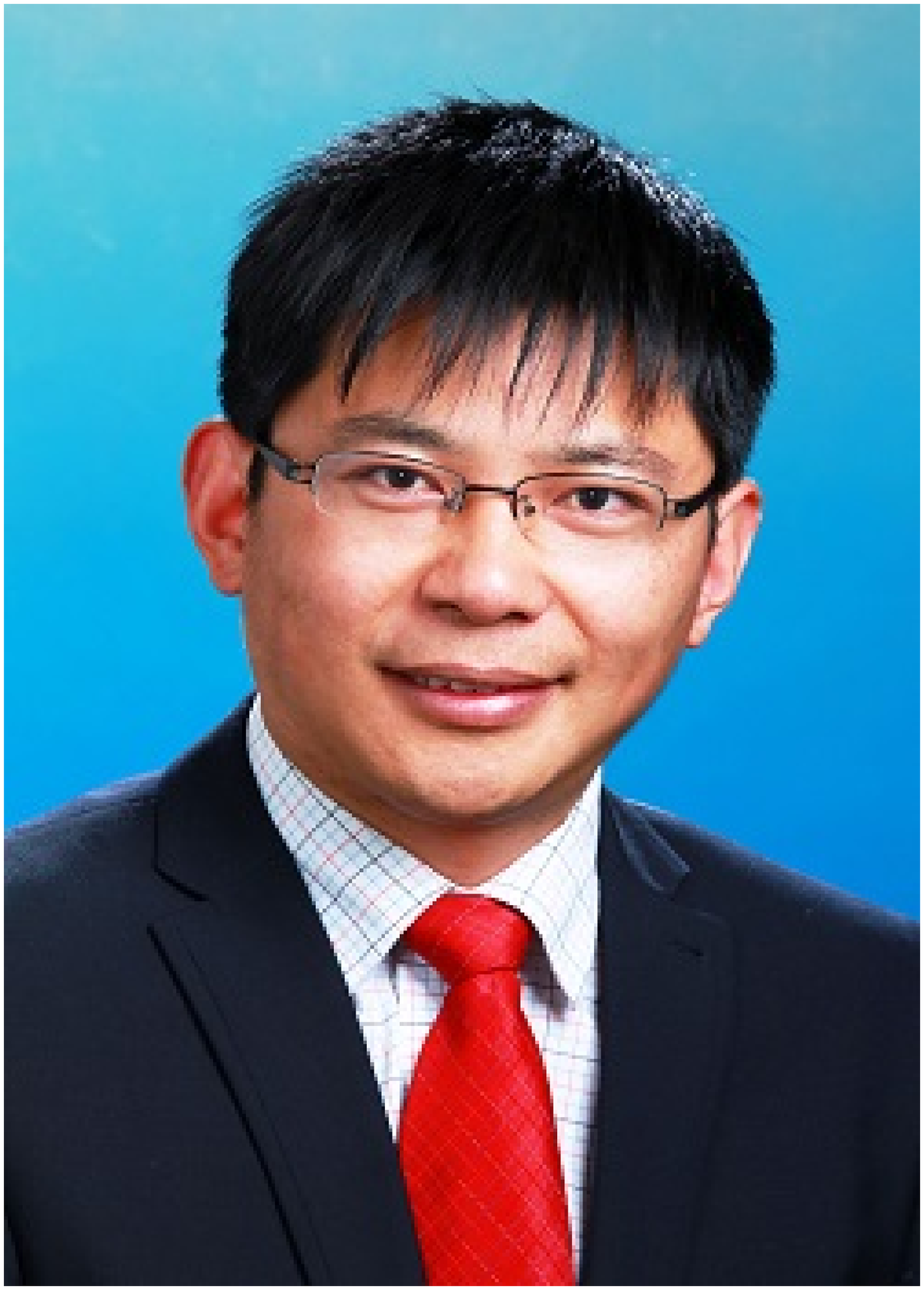}}]{Dajun Du}
received the B.Sc. and M.Sc. degrees all from the Zhengzhou University, China in 2002 and 2005, respectively, and his Ph. D. degree in control theory and control engineering from Shanghai University in 2010. From September 2008 to September 2009, he was a visiting PHD student at Queen’s University Belfast, UK. From April 2011 to August 2012, he was a Research Fellow at Queen’s University Belfast, UK. He is currently a  professor in Shanghai University. His main research interests include system modelling and identification and networked control systems.
\end{IEEEbiography}

\begin{IEEEbiography}[{\includegraphics[width=1in,height=1.25in,clip,keepaspectratio]{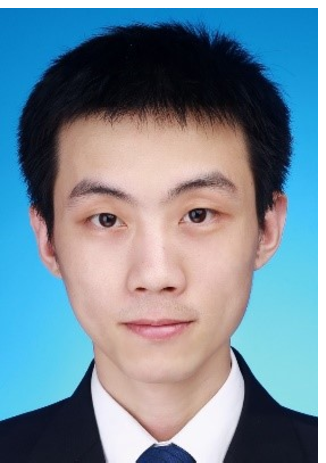}}]{Changda Zhang}
received the B.Sc. and M.Sc. degrees from Shanghai University, Shanghai, China in 2016 and 2019, respectively. He is currently pursuing the Ph. D. degree in Shanghai University. His main research interests include secure control for networked control systems.
\end{IEEEbiography}

\begin{IEEEbiography}[{\includegraphics[width=1in,height=1.25in,clip,keepaspectratio]{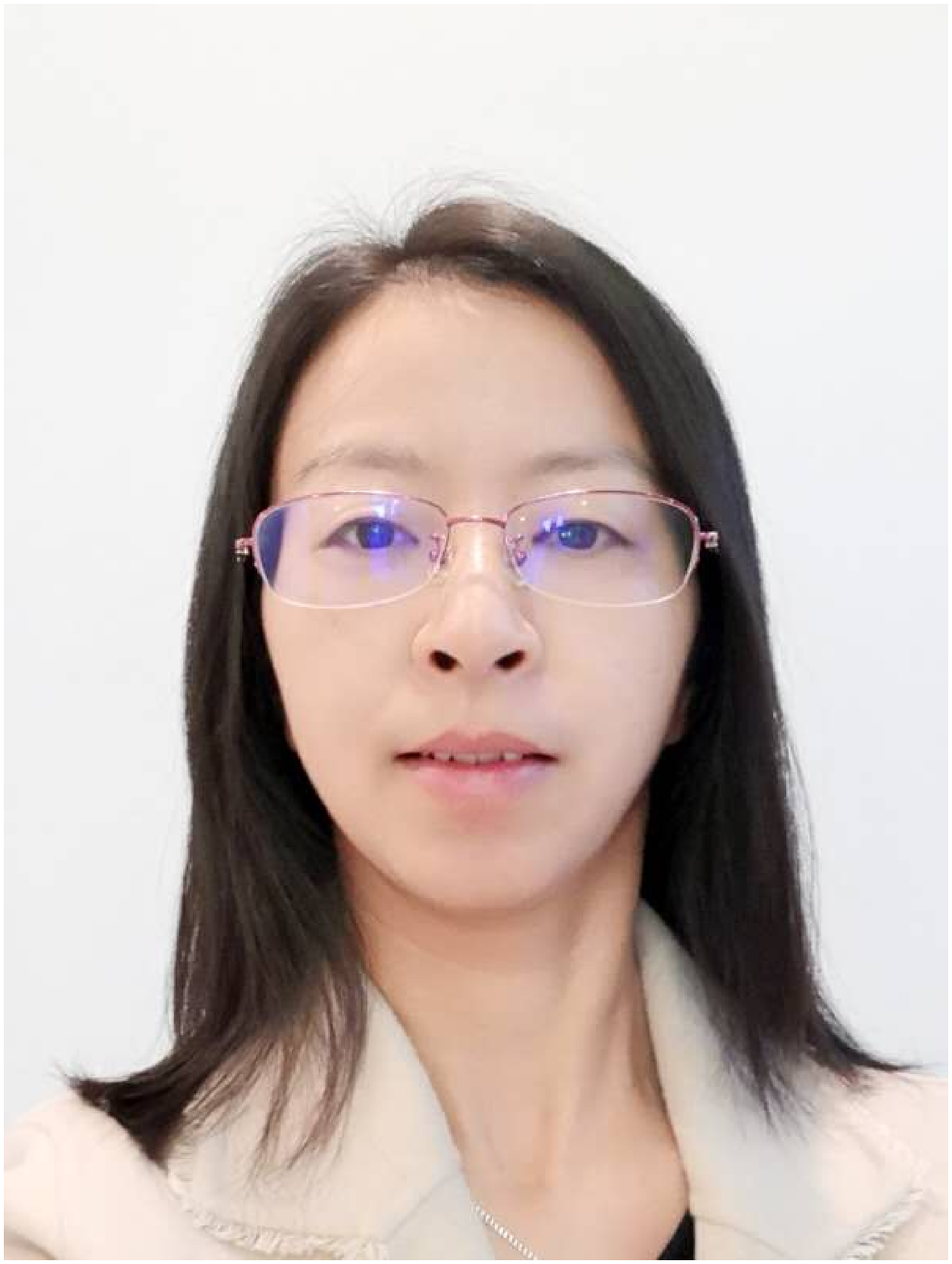}}]{Xue Li}
received the B.Sc. and M.Sc. degrees all from the Zhengzhou University, China in 2002 and 2006, respectively, and her Ph. D. degree in control theory and control engineering from Shanghai University in 2009. She is currently a professor in Shanghai University. Her main research interests include security control and performance assessment of smart grids.
\end{IEEEbiography}

\begin{IEEEbiography}[{\includegraphics[width=1in,height=1.25in,clip,keepaspectratio]{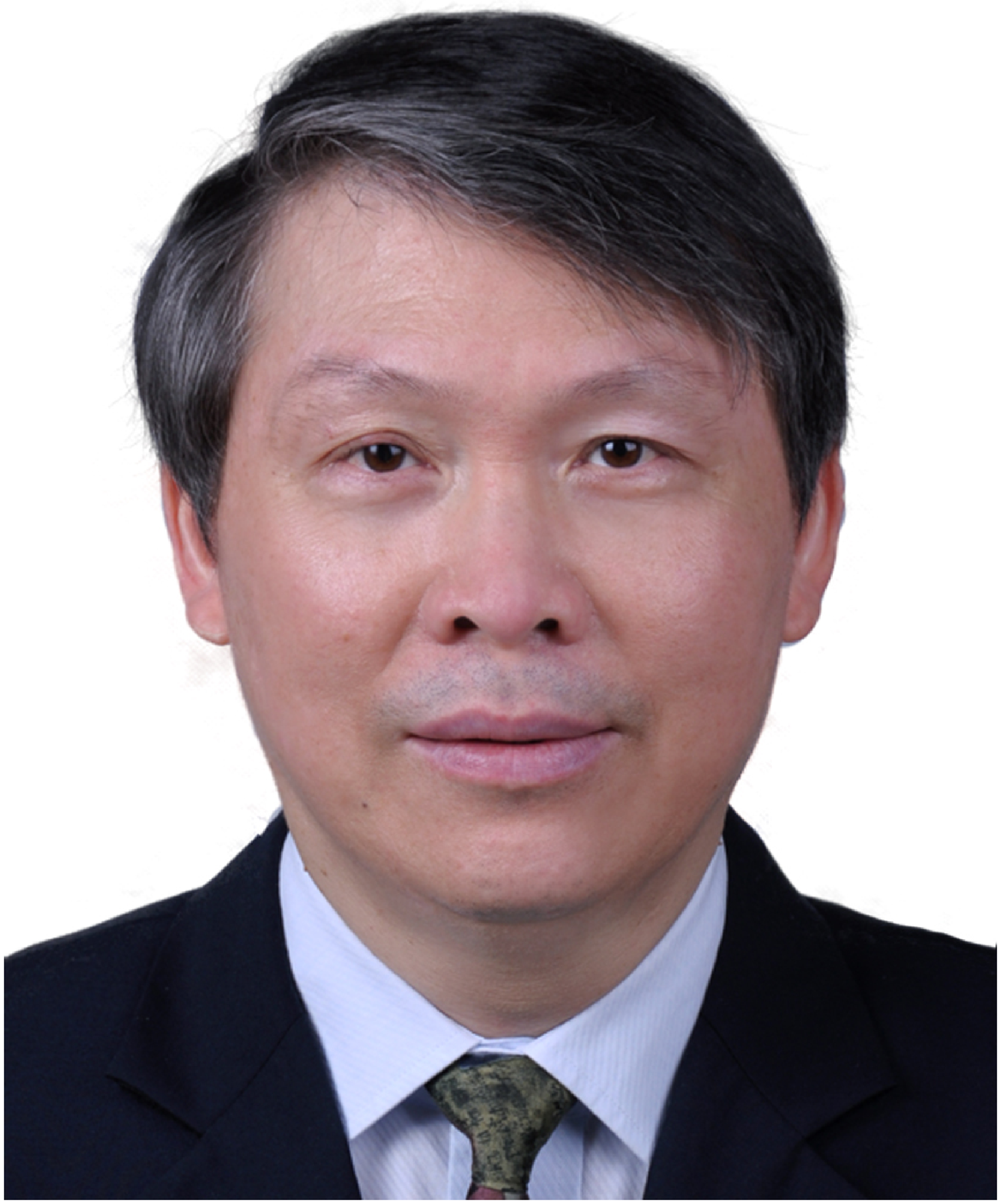}}]{Minrui Fei}
received his B.S. and M.S. degrees in Industrial Automation from the Shanghai University of Technology in 1984 and 1992, respectively, and his PhD degree in Control Theory and Control Engineering from Shanghai University in 1997. Since 1998, he has been a full professor at Shanghai University. He is Chairman of Embedded Instrument and System Sub-society, and Standing Director of China Instrument \& Control Society; Chairman of Life System Modeling and Simulation Sub-society, Vice-chairman of Intelligent Control and Intelligent Management Sub-society, and Director of Chinese Artificial Intelligence Association. His research interests are in the areas of networked control systems, intelligent control, complex system modeling, hybrid network systems, and field control systems.
\end{IEEEbiography}

\begin{IEEEbiography}[{\includegraphics[width=1in,height=1.25in,clip,keepaspectratio]{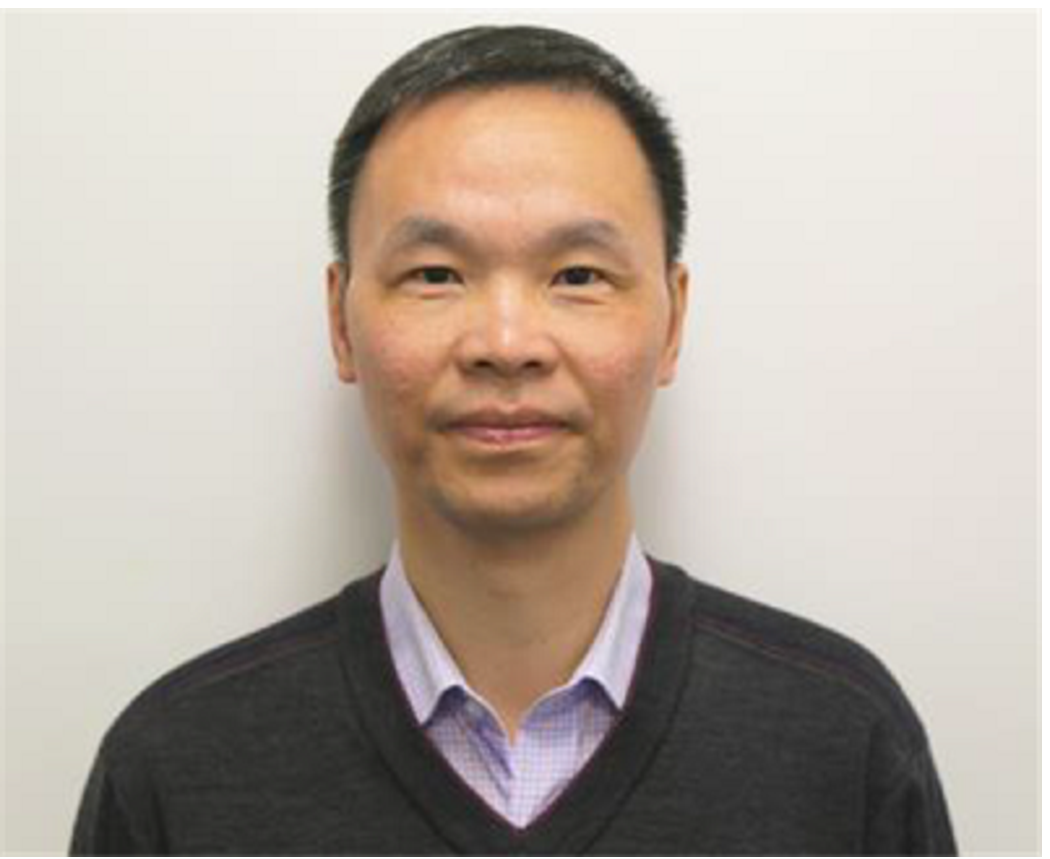}}]{Huiyu Zhou}
received his BEng degree in radio technology from Huazhong University of Science and Technology, China, a MSc degree in biomedical engineering from University of Dundee, U.K., and a Doctorate of Philosophy degree in computer vision from Heriot-Watt University, Edinburgh, U.K. He is currently a full Professor with School of Computing and Mathematical Sciences, University of Leicester, U.K. He has authored or co-authored over 350 peer-reviewed papers in the field. His research work has been or is being supported by UK EPSRC, MRC, AHRC, ESRC, EU ICT, Royal Society, Innovate UK, Leverhulme Trust, Invest NI, Puffin Trust, Alzheimer Research (UK) and industry.
\end{IEEEbiography}

\end{document}